\documentclass[a4paper, 10pt, onecolumn]{article}      % A4, 1 column
\usepackage[top=1in, bottom=1.25in, left=1.25in, right=1.25in]{geometry} %margins article
\usepackage[T1]{fontenc}
\usepackage[font=small,labelfont=bf,tableposition=top]{caption}

\usepackage{xfrac} %diagonal line at fractions

\DeclareCaptionLabelFormat{andtable}{#1~#2  \&  \tablename~\thetable}

\usepackage{floatrow}
\newfloatcommand{capbtabbox}{table}[][\FBwidth]
\usepackage{blindtext}

\usepackage{epsfig} % for postscript graphics files
\usepackage{epstopdf}
\usepackage{gensymb} %degree celsius symbol, by using \degree
\usepackage{graphicx}     %Display .png-pictures
\usepackage{amsmath}	%Useful for `gathered' and probably also for gather and align.

\usepackage{ulem}

\usepackage{breqn} %automatic line break equations

\usepackage{tabularx}

\usepackage{float}
\usepackage{subfig}
\usepackage{graphicx}

\allowdisplaybreaks		%With this command, page breaks within sets of equations become possible.

\usepackage{notoccite}

\title{\LARGE \bf
Long-term microstructural evolution of tungsten under heat and neutron loads}

\author{A. Mannheim, J.A.W. van Dommelen and M.G.D. Geers}% <-this % stops a space
\author{A. Mannheim, J.A.W. van Dommelen\footnote{Correspondence to J. A. W. van Dommelen. Electronic mail: j.a.w.v.dommelen@tue.nl}\hspace{0.1cm}  and M.G.D. Geers}

\date{Mechanics of Materials, Mechanical Engineering, Eindhoven University of Technology, P.O. Box 513, 5600 MB Eindhoven, The Netherlands \vspace{0.4cm}\\ \today}

\providecommand{\e}[1]{\ensuremath{\times 10^{#1}}}
\begin{document}

\maketitle
\thispagestyle{empty}

%%%%%%%%%%%%%%%%%%%%%%%%%%%%%%%%%%%%%%%%%%%%%%%%%%%%%%%%%%%%%%%%%%%%%%%%%%%%%%%%
\begin{abstract}
In nuclear fusion reactors, tungsten will be exposed to high neutron loads at high temperatures (>900 \textdegree C). The evolution and degradation of the mechanical properties under these conditions is uncertain and therefore constitutes a major risk. Here, the microstructural evolution of tungsten under combined heat and neutron loads is studied, using a multi-scale approach incorporating clusters dynamics and a mean-field recrystallization model. The mean-field recrystallization model contains both nucleation in the bulk and at the grain boundaries. The cluster dynamics model includes the incorporation of loops in the dynamics of the dislocation network as a mechanism. The effects of bulk nucleation on the microstructural evolution are explored. The simulations predict a cyclically occuring neutron-induced recrystallization at all studied temperatures. Furthermore, the evolution of the irradiation hardening during neutron-induced recrystallization is assessed from the simulated microstructures.

%Moreover, both bulk nucleation and necklace nucleation are included in the mean-field recrystallization model. 
%In this paper, more quantitative results for neutron-induced recrystallization are discussed. Loop unfaulting is included this time. The consequences of the microstructural changes on the thermal conductivity and yield strength of tungsten are discussed. The effects of nucleation in the bulk on the microstructural evolution are explored.
\end{abstract}

%%%%%%%%%%%%%%%%%%%%%%%%%%%%%%%%%%%%%%%%%%%%%%%%%%%%%%%%%%%%%%%%%%%%%%%%%%%%%%%%
\section{INTRODUCTION}

Nuclear fusion is a potential clean and cheap way to generate energy, making our society less dependent on oil, gas and coals. The most promising design for a large scale fusion reactor is the tokamak. Whereas the principle for fusion seems simple, the temperature required to achieve fusion, is extremely high (>150 million \textdegree C). The plasma-facing part of the divertor will receive high heat loads (10 MW/m$^2$ for DEMO \cite{Bolt2004}) and on top of that, plasma ion loads (peak loads of 10$^{24}$ ions/m$^3$s \cite{Bolt2004}) and neutron loads (15 dpa (displacements per atom) for 5 year of operation of DEMO \cite{Bolt2004}). These severe loading conditions need to withstood during its economic lifetime, which is at least two years before these critical parts can be replaced. \\
\\
The component in the wall that receives the highest heat loads, and that will thus reach the highest temperatures, is the divertor. The plasma-facing parts of this component will be made out of tungsten for ITER and DEMO, because of its favourable thermo-mechanical physical properties at high temperatures, its low activation levels and low tritium retention. Despite these qualities, it is not known how long tungsten can withstand these combined heat, neutron and ion loads. \\
\\
So far, one aims to keep the temperature of the divertor in between the DBTT and the recrystallization temperature, to prevent both brittle fracture and the loss of the pre-existing microstructure. However, as argued in \cite{Vaidya1983,Mannheim2018}, the combination of a continuous neutron load and a high temperature makes recrystallization and grain growth \cite{Kaoumi2008} hard to avoid. These processes cause a radical change in the materials microstructure and thus also in its properties, which makes the understanding of these processes critical. \\
\\
Neutron irradiation of tungsten leads to displacement damage and transmutations (to e.g. Re, Os, Ta) as well as gas production (H, He). The fusion neutron energy spectrum consists largely of 14 MeV neutrons. These fast neutrons mostly create large displacement cascades, in contrast to fission neutrons, which are more often thermal, producing relatively more transmutations. The evolution of the material structure, and thus of the material properties, is expected to be radically different under the fusion neutron spectrum than under fission neutrons. No experimental setups currently exist with a sufficient neutron flux for fusion neutron testing. Therefore, models for the evolution of material properties under fusion neutron irradiation are essential.\\
\\
The displacement cascades in tungsten have been observed with TEM to result in vacancies, self-interstitial atoms (SIAs) and clusters of these, which may form prismatic loops of interstitials and vacancies and voids \cite{Ferroni2015}. These interact with the pre-existing dislocation network, with the grain boundaries and with the material impurities. The interactions between the different types of defects are complex: defects may be trapped and impurities may stabilize or promote certain types of defects. Furthermore, the self-interstitial clusters are notorious for their large mobility in a certain direction. Object kinetic Monte Carlo (OKMC) models as well as stochastic cluster dynamics models become increasingly more complete in describing the damage evolution, rendering them more accurate in predicting the defects that are observed experimentally \cite{Castin2018, Huang2018}. However, OKMC models are only able to describe the microstructural evolution in a volume that is several orders of magnitude smaller than the size of a single grain. Hence, they cannot be used to describe the evolution of a polycrystalline structure. Although less accurate, cluster dynamics (CD) models are computationally more efficient and better suited for use in polycrystalline modelling.\\
\\
A first step towards a multi-scale model of the microstructural evolution under combined heat and neutron loads was made previously in \cite{Mannheim2018}. With this model it is possible to study the competition between the various physical mechanisms and to assess their effects on the resulting grain size distribution and defect densities. In this model, the interaction between dislocations, vacancies, self-interstitial atoms and their clusters, modelled using CD modelling \cite{Li2012}, determines the defect concentrations. Based on these concentrations, the driving forces for grain growth and nucleation are determined, using a mean-field recrystallization model \cite{Bernard2011}. \\
\\
In this paper,  several important extensions to the modelling framework \cite{Mannheim2018} for irradiation-induced recrystallization are made. The mechanism of loop incorporation into the dislocation network has been reported in \cite{Was2007, Zinkle2012, Patra2016} to occur in both bcc and fcc metals and was observed for tungsten in the form of coalescence of loops and dislocation network formation under certain conditions \cite{Ferroni2015}.  This mechanism is included in the present model and its effect on the microstructural evolution is investigated. Furthermore, nucleation at grain interiors (bulk nucleation) due to a high defect density as a result of neutron irradiation is incorporated and the effect of the amount of bulk nucleation on the evolution of the polycrystalline structure is explored. Lastly, the evolution of the hardness during neutron-induced recrystallization is predicted, based on the microstructural evolution.

\section{Method}

The microstructural evolution of polycrystalline tungsten under irradiation damage is modelled using a multi-scale approach. As a basis, the model as described in \cite{Mannheim2018} is exploited, whereby several extensions are made in order to obtain a more realistic description of the microstructural behaviour. \\
\\
The microstructural evolution that results from grain growth and nucleation, processes that occur at higher temperatures, are described using a mean-field model \cite{Bernard2011}. These processes are driven by the stored lattice energy that accumulates in the grains as a result of displacement cascade damage under neutron irradiation. The displacement damage consists of vacancies, vacancy clusters, self-interstitial atoms and their clusters and dislocations. The evolution of the defect concentrations and dislocation density inside each grain is described using cluster dynamics \cite{Li2012, Mannheim2018}. The coupled model describes the evolution of the grain sizes $r$, the defect concentrations $C$ and dislocation densities $\rho$ for a set of grains that are representative for the microstructure.\\
\\
In this section, the recrystallization model and the cluster dynamics model are described, as well as the solution procedure for the coupled model. Details on the implementation are given in Appendix~\ref{sec:appimp}. 

\subsection{Recrystallization}
\begin{figure}[ht!]
	\begin{minipage}[c]{.48\textwidth}
		\vspace{0.07\textheight}
		\subfloat[]{\includegraphics[height=0.15\textheight]{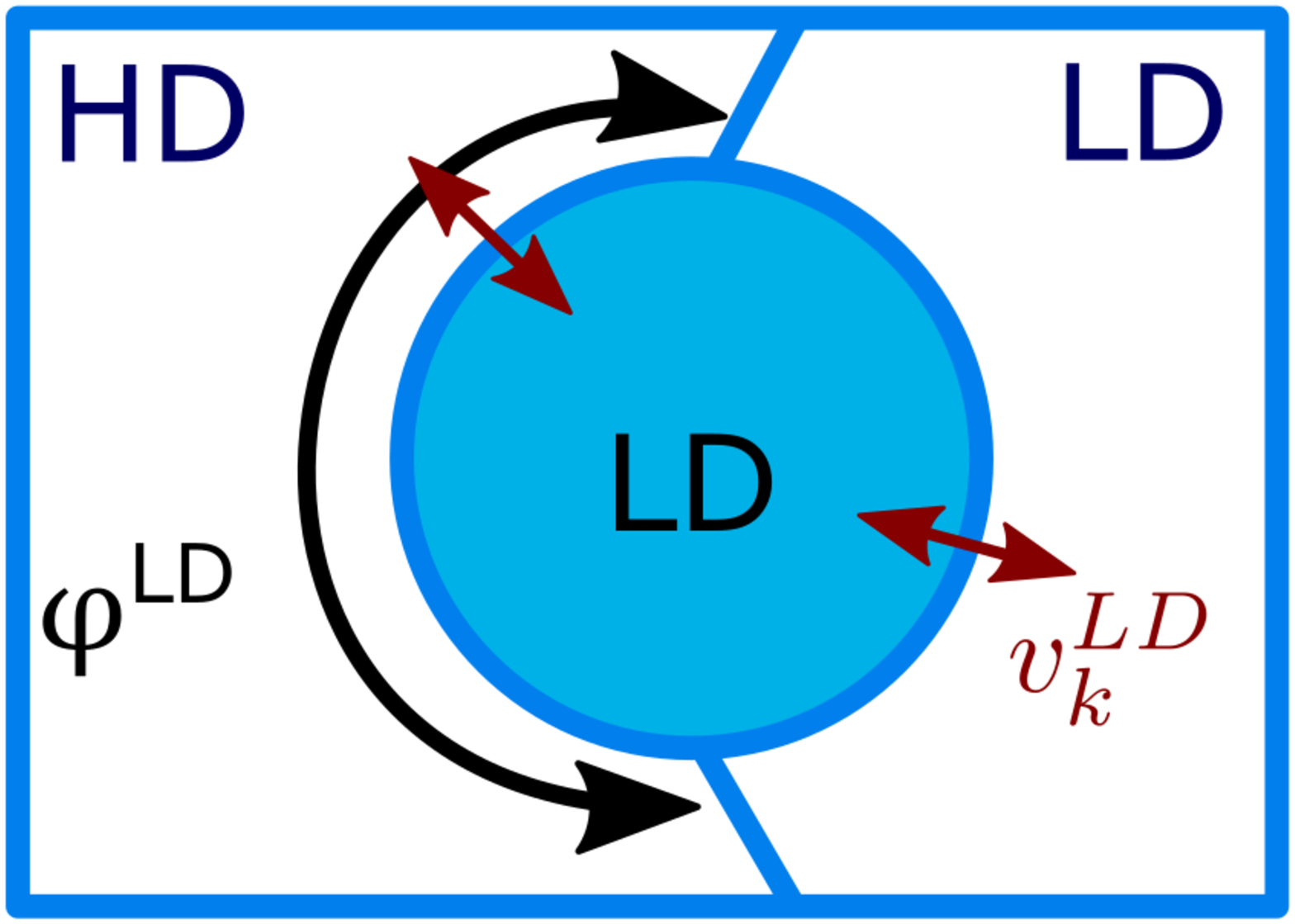}}
	\end{minipage}
	\begin{minipage}[c]{.48\textwidth}
		\hspace{1cm}
		\subfloat[]{\includegraphics[height=0.22\textheight]{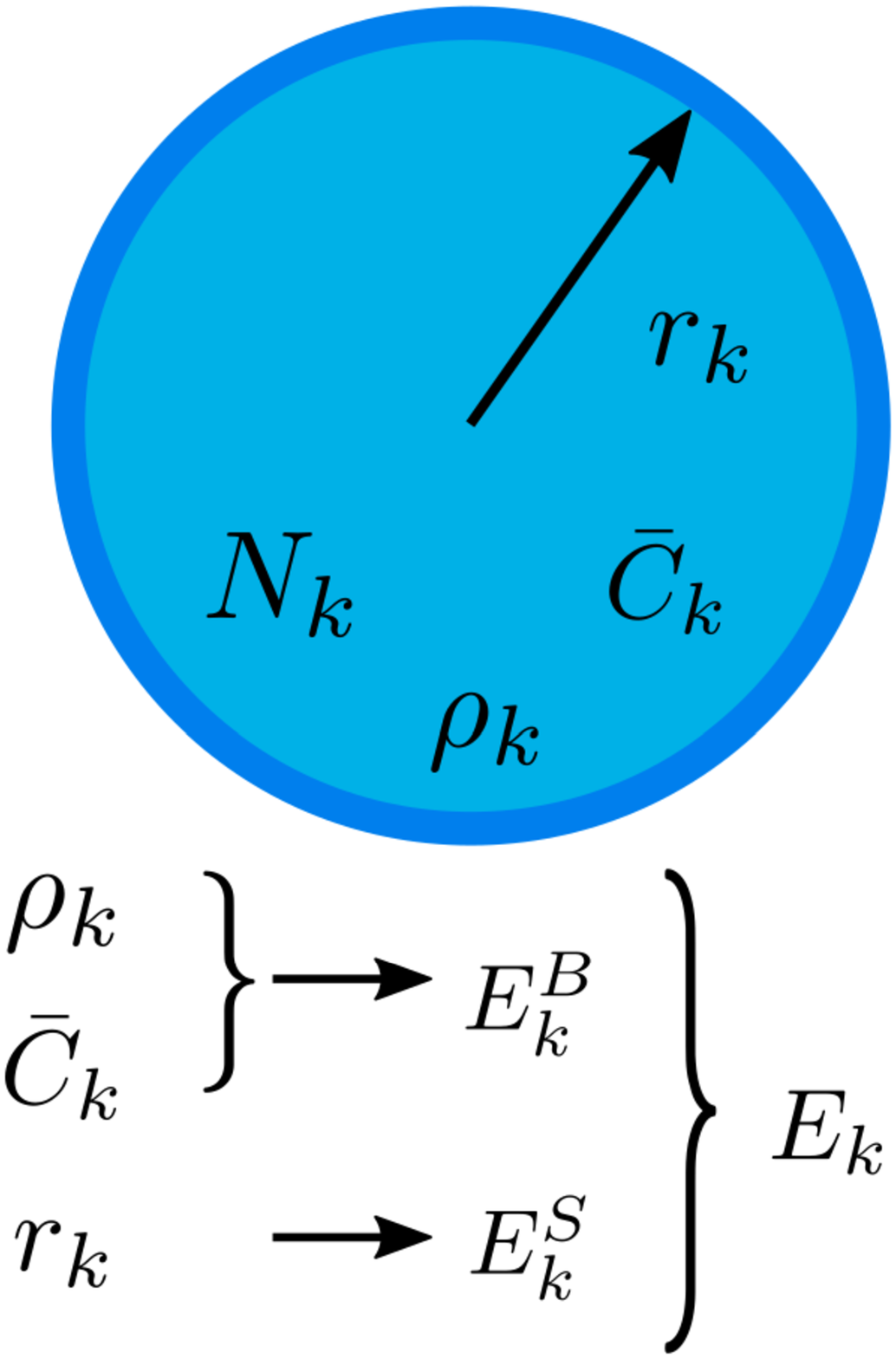}}
	\end{minipage}
\caption[]{Sketch of the mean-field model, with (a) a representative grain from the LD-medium interacts with the surrounding LD- and HD-media, and (b) a representative grain, including its properties.}
\label{fig:mf}
\end{figure}

In the two-media mean field model \cite{Bernard2011,Mannheim2018}, the microstructure is represented by a set of grains that are distributed over a homogeneous equivalent medium with low defect density (LD-HEM) and a medium with high defect density (HD-HEM), see Figure~\ref{fig:mf}a. \\ \\
Each representative grain $k$ is assumed to be spherical and has as evolving properties: a radius $r_k$, a number of grains $N_k$, defect densities $\rho_k$ or defect concentrations $C_k$, see Figure~\ref{fig:mf}b. At a given moment in time, the stored bulk (i.e. excluding boundaries/surfaces) energy density $E^B_k$ of grain $k$ is calculated using the defect formation energies (see Appendix~\ref{app:rc}), defect concentrations and the configurational entropy (Appendix~\ref{app:ent}), as illustrated in Figure~\ref{fig:mf}b. The surface energy density $E^S_k$ of a grain follows from the grain boundary surface energy $\gamma_b$ and the grain radius $r_k$. When for a grain, $E^B_k$ exceeds the energy density threshold $E_T$, this grain is allocated to the HD-HEM. The grains interact with both their averaged surroundings (i.e. the HEMs). The stored energy in the material drives grain growth and nucleation, as a result of which the grain properties (grain size and defect concentrations) evolve.

\subsubsection{Grain growth}
During grain growth, a grain boundary segment that is shared between grain $k$ and a HEM, moves towards the HEM with velocity 
\begin{equation}
v^{HEM}_k=m \bigg(E^{HEM}-E_k\bigg) = m \bigg(E^{B,HEM} - E^B_{k} + \frac{3 \gamma_b}{2} ( \frac{1}{r^{HEM}} - \frac{1}{r_k})\bigg),
\end{equation}where $m$ is the temperature-dependent grain boundary mobility, $E^{HEM}$ is the volume average of the stored energy densties of the grains in the HEM, $E_k$ is the total stored energy density of grain $k$, $E^{B,HEM}$ and $E^{B}_k$ are the bulk parts (i.e. excluding boundaries/ surfaces) of the stored energy densities of the HEM and grain $k$, and $r^{HEM}$ and $r_k$ are the volume averaged radius of the HEM and the radius of grain $k$, respectively \cite{Mannheim2018}. The velocities of both segments of the grain boundary (one for each HEM, see Figure~\ref{fig:mf}a) lead to a volume change, and therefore to a new radius for the grain. The volume that is added to grains that are subject to growth, has thermal equilibrium defect concentrations $C^{eq}$ and a thermal equilibrium dislocation density $\rho^{eq}$. For a growing grain, the average defect concentrations inside the grain thus diminish \cite{Mannheim2018}.\\
\\
The fraction of the total grain boundary surface area of an LD-grain that is shared with the HD-HEM is taken as $\phi^{LD}=1-(f^{LD})^{p}$ (see Figure~\ref{fig:mf}a), with $f^{LD}$ the total volume fraction of the microstructure that resides in the LD-HEM and $p = 2/3$ for necklace-type nucleation \cite{Bernard2011}. For HD-grains, the fraction of the grain boundary surface that is shared with the LD-HEM is denoted as $\phi^{HD}$ and is determined each time such that the volume transfers between the LD-medium and the HD-medium are equal \cite{Bernard2011}. The maximum of $\phi^{HD}$ is limited to 1, and $\phi^{LD}$ may be adjusted accordingly \cite{Bernard2011}, see Appendix~\ref{sec:appimp}.\\
\\
The mobility of the grain boundaries $m$ is strongly temperature dependent. The Turnbull estimate \cite{Cram2009} is used as a basis:
\begin{equation}
m(T) 	= K_m \frac{ \delta V_{at}}{b^2 RT} D^{GB}_0 \text{exp}\bigg( \frac{-Q^{GB}}{k_B T} \bigg).
\end{equation} $K_m$ was characterized based on measurements on static recrystallization of tungsten in \cite{Mannheim2018}, $Q^{GB}$ is the activation energy for diffusion of tungsten along grain boundaries, $\delta$ is the thickness of the grain boundaries, $V_{at}$ is the atomic volume, $R$ is the gas constant, $T$ is the temperature, $D^{GB}_0$ is the self-diffusivity of tungsten along the grain boundaries and $k_B$ is the Boltzmann constant. The parameter values are given in Appendix~\ref{app:par}. The implementation of the grain growth model is further described in section~\ref{par:gg}.

\subsubsection{Nucleation} \label{sec:nuc}
\begin{figure}[ht!]
\subfloat[]{\includegraphics[width=0.4 \textwidth]{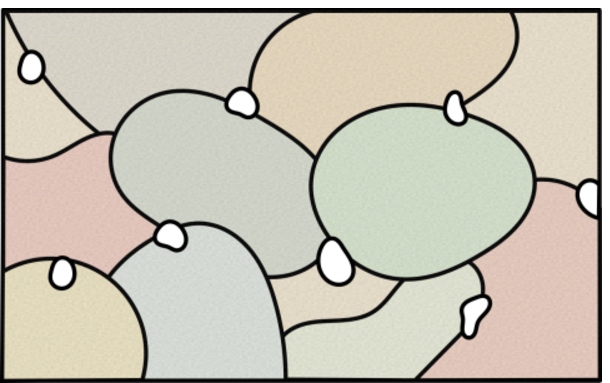}}
\hspace{0.5 cm}
\subfloat[]{\includegraphics[width=0.4 \textwidth]{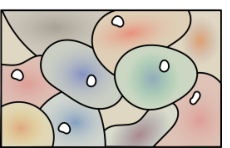}}
\caption[]{(a) Sketch of necklace-type nucleation in a polycrystal. (b) Nucleation taking place in the grain interiors as a result of the higher stored energy at that location (bulk nucleation).}
\label{fig:nuc}
\end{figure}

The formation of stable nuclei is driven by a reduction in the Gibbs free energy $E$. Often, nucleation occurs more easily at the grain boundaries (necklace-type nucleation, see Figure~\ref{fig:nuc}a), where the activation barrier is reduced. EBSD-images taken during static recrystallization of pre-deformed tungsten indicate that necklace-type nucleation occurs in tungsten as well \cite{Lopez2015}. However, due to displacement cascades induced by neutrons, a high lattice stored energy may accumulate inside the grains, which triggers an additional nucleation mechanism, inside the grains (here referred to as bulk nucleation, see Figure~\ref{fig:nuc}b). 

It has been suggested that under irradiation conditions, nucleation could be initiated at prismatic dislocation loops \cite{Rest2004}. Here, both types of nucleation are modelled. The total rates for necklace nucleation $\dot{N}^A$ and bulk nucleation $\dot{N}^V$ in the microstructure are described by:
\begin{align}
\dot{N}^A=K_N^A A_{nuc} \exp{\bigg(\frac{-E_{act}^A}{k_B T}\bigg)} \exp{\bigg(\frac{-Q^{GB}}{k_B T}\bigg)}, \\
\dot{N}^B=K^B_N V_{nuc} \exp{\bigg(\frac{-E^B_{act}}{k_B T}\bigg)} \exp{\bigg(\frac{-Q^{GB}}{k_B T}\bigg)}.
\end{align}Here, $K_N^A$ and $K_N^B$ are the nucleation rate constants, $A_{nuc}$ and $V_{nuc}$ are the total available grain boundary area and nucleation volume respectively, and $E_{act}^A$ and $E_{act}^V$ are the activation energies for the formation of a stable grain. The activation energy and the radius of a stable nucleus $r_{nuc}$ are determined by solving $\frac{\partial \Delta E}{\partial t}<0$ (where $\Delta E$ denotes the Gibbs free energy difference) under the condition that the velocity of the grain boundary of the nucleus is positive, $v^{nuc}_{GB}>0$ \cite{Mannheim2018}. The expressions for $\frac{\partial \Delta E}{\partial t}$ are given in section~\ref{sec:imp}.\\
\\
During bulk nucleation, a new grain nucleates and grows completely at the cost of the HD grain interior (Figure~\ref{fig:nuc}b). Therefore, for a nucleated grain of radius $r$, the Gibbs free energy changes as: $\Delta E^B=\frac{1}{K^B_a} \bigg[ -\frac{4 \pi r^3}{3} (  E^{B,HD} - E^B_{0}) + 4 \pi r^2 \gamma_b \bigg]$, where $K_a^B$ is a parameter for the reduced activation energy during bulk nucleation, $E^{B,HD}$ is the volume average of the bulk (i.e. excluding boundaries/surfaces) stored energy density for the grains in the HD-medium, $E^B_0$ is the bulk stored energy density for a grain with equilibrium defect concentrations and $\gamma_b$ is the grain boundary surface energy. The grain boundary velocity of the newly nucleated grain that is volumetrically growing, is taken as $\frac{dr}{dt}=m(E^{HD} - E_0)$, with $E^{HD}$ the volume average of the total energy density of the HD-grains and $E_0$ the total stored energy density of the nucleated grain, including its grain boundary energy density. The grains that nucleate in the bulk, are referred to as HD daughter grains, as they initially only grow with respect to their HD parent grains. All the other grains in the microstructure are referred to as regular grains. In section~\ref{sec:imp}, details about the treatment of HD daughter grains can be found.\\
\\
In the case of necklace nucleation, the nucleus immediately can have both LD-grains and HD-grains as neighbours (Figure~\ref{fig:nuc}a). It is assumed that during necklace nucleation, $3\pi r^2$ of extra grain boundary area is formed \cite{Mannheim2018}. The associated Gibbs free energy change is $\Delta E^A=\frac{1}{K_a^A} \bigg[- \frac{4 \pi r^3}{3} \bigg( E^B - E^B_{0} \bigg) + 3 \pi r^2 \gamma_b \bigg]$, where $K_a^A$ is a parameter for the reduced activation energy at grain boundaries. In this case, the grain boundary velocity right after nucleation is $\frac{dr}{dt} = m(E-E_0)$, where $E$ is the volume average of the stored energy density of all the grains in the microstructure. The surface area $A_{nuc}$ available for necklace nucleation consists of all the HD/LD grain boundaries (excluding the grain boundaries of the HD daughter grains) and all HD/HD grain boundaries.

\subsection{Cluster dynamics}
Highly energetic neutrons can create cascades of displaced atoms, which result in the creation of vacancies ($V_1$), self-interstitial atoms ($I_1$) and their clusters $V_n$ and $I_n$ in the lattice with $n$ the size of the cluster. The concentrations $C_{I_n}$ and $C_{V_n}$ of all these different defects are the degrees of freedom in this model. Clusters from size $n=1,\dots,N_{max}$ are considered. All the vacancy clusters are assumed to take the shape of spherical voids, following \cite{Li2012}, and all the SIA-clusters of more than 5 SIAs are assumed to form prismatic dislocation loops (a two-dimensional disk, with a dislocation line on the edge of this disk) \cite{Li2012}. 

\begin{figure}[ht!]
\includegraphics[width=0.6\textwidth]{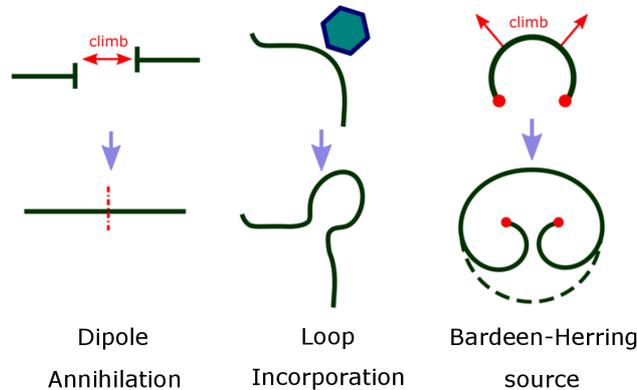}
\caption[]{Three mechanisms that lead to a change in the dislocation density are shown schematically: dipole annihilation, loop incorporation and a Bardeen-Herring source.}
\label{fig:dismech}
\end{figure}

All displacement defects are considered to be immobile, except for single vacancies and single interstitials. The defects can change in size by absorption/emission of a mobile defect of the same type or of the opposite type (e.g. $I_3 + V \rightarrow I_2$, or $I_6 \rightarrow I_5 + I$). The mobile defects can also be absorbed at grain boundaries and dislocations, that act as a sink for them \cite{Li2012}. The absorption at dislocations entails dislocation climb \cite{Stoller1990}. The climb motion can lead to dipole annihilation of two dislocations of the opposite type \cite{Stoller1990}. When a dislocation segment is pinned at two points, climb of the segment can result  in the creation of extra dislocation length (a Bardeen-Herring source) \cite{Jourdan2015}. When the prismatic loops (which are considered immobile) absorb interstitials or emit vacancies, they may grow, or  when they get in contact with other dislocations (either another prismatic loop or another dislocation line/segment), they become part of the dislocation network. This is modelled using the method of Jourdan \cite{Jourdan2015}, which was originally applied for faulted loops in austenitic stainless steels. Under irradiation conditions, dislocation loops first form and grow, subsequently unfault and become perfect loops. Thereafter, they interact and become part of the dislocation network \cite{Zinkle2012}. In bcc metals, prismatic loops unfault at very small sizes, at least for Fe \cite{Zinkle2012}. \\
\\
Note that even though a cluster dynamics model does not allow for direct dipole annihilation of prismatic dislocation loops, by including loop incorporation into the dislocation network density model, it becomes possible for the interstitial dislocation loops to annihilate. In Figure~\ref{fig:dismech}, the considered mechanisms for a change in dislocation density are shown schematically.\\
% and note also that in bcc metals, prismatic loops are formed as faulted loops but unfault into perfect loops at very small sizes, at least for Fe \cite{Someone}.
\\
The concentrations of the interstitial clusters $C_{I_n}$ and the vacancy clusters $C_{V_n}$ of size $n$ and the dislocation density $\rho$ inside an irradiated grain follow from this set of equations: 
\begin{alignat}{2}
&\frac{d}{dt} \bigg[ C_{I_n} \bigg]  	&&  	=  G_{I_n} + (1-f_n) J^I_{n-1,n} + J^I_{n+1,n} - [J^I_{n,n-1} + (1-f_{n+1}) J^I_{n,n+1}]  - L_{I_n} C_{I_n} , \label{eq:gen1}\\
&\frac{d}{dt}\bigg[ C_{V_n} \bigg]  	&&  	= G_{V_n} +  J^V_{n-1,n} + J^V_{n+1,n} - [J^V_{n,n-1} + J^V_{n,n+1}]  - L_{V_n} C_{V_n} , \\
&\frac{d}{dt}\bigg[ \rho \bigg]    	&&      = 2 \pi v_{cl} S_{BH}  +\frac{d \rho_{IC}}{dt}  - \rho \tau_{cl}^{-1} . 
\label{eq:gen3}
\end{alignat}
In the first two equations, the concentrations of the interstitial and vacancy clusters of size $n$ change over time by defect generation (sources $G_{I_n}$ and $G_{V_n}$), by transport of mobile defects to/from the defect clusters of size $n$ providing fluxes $J$ between cluster size $n$ and the adjacent cluster sizes $n-1$ or $n+1$ (e.g. $J_{n-1,n}$ is the flux from $n-1$ to $n$), and by annihilation of mobile defects at dislocation and grain boundary sinks of total strengths $L_{I_n}$ and $L_{V_n}$. For the interstitial clusters, a fraction $f_n$ of the loops of size $n$ that is about to grow, is incorporated in the dislocation network instead \cite{Jourdan2015}. The third equation expresses the dislocation density evolution: more dislocations are produced by Bardeen-Herring sources based on the climb velocity $v_{cl}$ and the source density $S_{BH}$, as well as by the incorporation of the prismatic loops $d \rho_{IC}/dt$. Dislocations annihilate by dipole annihilation, based on the average lifetime $\tau_{cl}=d_{cl}/v_{cl}$ of dislocation segments, which depends on the climb velocity $v_{cl}$ and on the average distance between the dislocations $d_{cl}$. See Appendix~\ref{app:dis} for more details. The transport fluxes $J$ and fluxes to the sinks are specified in Appendix~\ref{app:cd}.\\
\\
The production rate of defect clusters induced by neutron irradiation is based on MD-simulations of displacement cascades (50 ps simulation time) \cite{Setyawan2015}. The majority of the neutron-induced defects vanish \cite{Nordlund2018} by recombination of vacancies with self-interstitial atoms during thermal aging of the cascade. CD-modelling does account for recombination, as long as spatial correlations do not play an important role. A sophisticated method for transferring MD-results for neutron cascade damage into the CD-model, bridging the time scales by the use of a kinetic Monte Carlo model, can be found in \cite{Jourdan2018}. Here, a simplified approach is taken: the power-law dependence found in \cite{Yi2015} is assumed to hold, and a parameter $\eta$ denotes the remaining defect fraction after prolonged annealing of the MD cascade.
The production rate for a defect cluster of size $n$ is then given by
\begin{align}
G_n = (1-f) \eta A_{\epsilon} / n^{S_{\epsilon}},
\label{eq:G}
\end{align}where $f$ is the total amount of vacancies and interstitials per atom, and with $\epsilon=I$ or $V$, based on \cite{Yi2015,Mannheim2018}. Here $A_{\epsilon}$ and $S_{\epsilon}$ are temperature dependent constants that are based on MD-results, see Table~\ref{tab:ownpowerlaw} in Appendix~\ref{app:par}. 

%\textcolor{blue}{The production rate of defect clusters by the neutron irradiation is based on MD-results.} 
%The production rate for a defect cluster of size $n$ is given by $G_n = D_{eff}(1-f)A_{\epsilon} / n^{S_{\epsilon}}$, \textcolor{blue}{where $n$ is the size of the defect cluster,} $f$ is the total amount of vacancies and interstitials per atom, and with $\epsilon=I$ or $V$, based on \cite{Yi2015,Mannheim2018}. Here $A_{\epsilon}$ and $S_{\epsilon}$ are temperature dependent constants that are based on MD-results, see Table~\ref{tab:ownpowerlaw} in Appendix~\ref{app:par}. The transport fluxes $J$ and fluxes to the sinks are specified in Appendix~\ref{app:cd}. \textcolor{blue}{A more thorough method for transferring MD-results on  neutron cascade damage into the CD-model, involving the use of kMC, is described in \cite{Jourdan2018}.}\\

\paragraph{Loop incorporation} The expression for the rate with which dislocation loops are incorporated in the pre-existing dislocation network (i.e. loop incorporation) is based on \cite{Jourdan2015} and is given by:
\begin{dmath}
\frac{d \rho_{IC}}{dt}= \sum_{n=5}^{N_{max}} 2 \pi r_n f_n (\rho_t) J^I_{n-1,n} = \sum_{n=5}^{N_{max}} 2 \pi r_n f_n (\rho_t) [\alpha_{n-1}^+ C_{I} C_{I_{n-1}}  + k^-_{I_{n-1}-V} C_{I_{n-1}}].
\end{dmath}Here 5 is the minimum size adapted for a SIA-cluster to form a prismatic loop (in analogy with \cite{Stoller1990}), $r_n$ is the radius of the prismatic loop, the terms between the brackets are the growth rates of the prismatic loops from size $n-1$ to size $n$ by absorption of a self-interstitial atom/emission of a vacancy (further specified in Appendix~\ref{app:rc}) and $f_n$ is the fraction of the growing prismatic loops of size $n$ that are incorporated, \textit{i.e.} touching other dislocations. This fraction depends on the total dislocation density $\rho_t$ (the sum of all the prismatic loop dislocation densities and the network dislocation density) \cite{Jourdan2015}: 
\begin{align}
f_n(\rho_t) & = 1- \frac{\text{erfc}\bigg({\frac{h_n-h_t}{\sqrt{2} \sigma_d}}\bigg)}{\text{erfc}{\bigg(\frac{h_{min}-h_t}{\sqrt{2} \sigma_d}}\bigg)}
\end{align}with $h_n=2 r_n $, $h_t = \frac{2}{\sqrt{\pi \rho_t}}$, $\sigma_d=0.6 h_t$ and $h_{min}=2 r_{min} = 2 r_{5}$.

\paragraph{Absorption/emission} The absorption rates of mobile defects at defect clusters are modelled as pure diffusion-limited \cite{Li2012}. For example, the absorption rate of an interstitial to a prismatic interstitial loop of size $n$ is given by $\alpha^+_n=2 \pi r^{cap}_{I_n} Z^I_{I_n} D_I$, with $Z^I_{I_n}$ the factor of preferential absorption of interstitials to dislocation loops, $D_I$ the diffusion coefficient for interstitials  and $r^{cap}_{I_n}$ is the capture radius of the prismatic loop \cite{Li2012}. For prismatic loops the capture radius is $r^{cap}_{I_n}=\sqrt{nV_{at}/\pi b}$, where $V_{at}$ is the atomic volume and $b$ the length of the Burgers vector \cite{Li2012}. Similar expressions are used for the other interactions of mobile defects and clusters, as listed in Appendix~\ref{app:rc}. For vacancy clusters, the capture radius is taken as $r^{cap}_{V_n}=(3nV_{at}/4 \pi)^{1/3} + \sqrt{3} a_0/4$ \cite{Li2012}, where $a_0$ is the lattice parameter. The rates of emission of mobile defects also depend on the binding energy of the defect to the cluster, e.g. $E^b_{I_n-I}$. The binding energies are calculated using the capillarity approximation, see Appendix~\ref{app:rc}.

\paragraph{Sinks for mobile defects} The sink strength $L_{\epsilon_n}$ (where $\epsilon$=$I$ or $V$, for interstitials and vacancies) is the sum of the grain boundary sink strength $k^+_{S+\epsilon}$ and the dislocation sink strength $k^+_{D+\epsilon}$. The dislocation sink strength is given by $k^+_{D+\epsilon} =\rho Z^{\epsilon}_D D_{\epsilon}$, where $Z_D^{\epsilon}$ is a bias factor to account for preferential absorption of interstitials. The grain boundary sink strength ($k^+_{S+\epsilon}$) depends on the sum of the sink strengths of all the other sinks within the grain (dislocations as well as defect clusters), and on the grain radius. See Appendix~\ref{app:rc} for the precise expressions. 

\paragraph{Simulation parameters} A maximum cluster size $N_{max}$ = 100 is used in the simulations, along with $\eta=1$ for the damage production rate. It was verified that, compared to $N_{max}$ = 1000, the evolution of the atomic defect fraction $f$ was in reasonable agreement for most damage rates (varying $\eta$ from 0.001 to 1) at each of the irradiation temperatures (1000-1300\textdegree C).

\subsection{Solution procedure}\label{sec:imp}
The two-level model is solved incrementally, as follows:
\begin{enumerate}
\item Determination of the time step for the next time increment $\Delta t = t_{i+1}-t_i$ (see below);
\item Defect evolution: application of the cluster dynamics model (using Equations \ref{eq:gen1} - \ref{eq:gen3}) results in updated defect concentrations for each representative grain. Parallel computation is used in this step;
\item Assignment of the grains to the HEMs: based on their bulk stored defect energy densities $E^B$, the grains are placed in the HD-set when their energy exceeds the energy density threshold $E_T$, (see below, HEM allocation). If HD daughter grains (see section~\ref{sec:nuc}) are placed in the HD-set, they are subsequently treated as regular grains;
\item Nucleation: both for necklace- and bulk-type nucleation, the nucleation rates ($\dot{N}^A$ and $\dot{N}^V$), the nucleus radii ($r_{nuc}^{A}$ and $r_{nuc}^V$) and the activation energies ($E_{act}^A$ and $E_{act}^V$) are determined, as well as the nucleation surface area ($A_{nuc}$) and nucleation volume ($V_{nuc}$), and new grains are defined. The new nuclei are either represented by a new LD-grain, or they are merged with an existing representative grain (see below, HEM allocation). The procedure for nucleation is described in more detail in Appendix~\ref{sec:appimp};
\item Grain growth: first, all grains except for the HD daughter grains interact with both HEMs. As a result, the grain sizes and defect densities change. If grains are vanishing, subincrements are used in this step. The use of subincrements is detailed in Appendix~\ref{sec:appimp}. Finally, the HD-daughter grains interact with the HD-HEM only, and the grain sizes and defect densities are updated accordingly.
\end{enumerate}

\paragraph{Time step}
The size of the time step from $t_i$ to $t_{i+1}$ is chosen in such a way that during a single increment, none of the grains $k$ (except grains that are already very small, $r<$0.3 $\mu $m) grow or shrink with more than 2\% of their volume, based on the volume changes $| \Delta V_k|/V_k$ determined in the previous time step $t_i$:
\begin{equation}
\Delta t_{i+1} < 0.02 \Delta t_i / [\max\limits_{k}(|\Delta V_k|/V_k)].
\end{equation}Moreover, the size of the time step is constrained to increase only slowly (2-5\% depending on the value) with respect to the previous time step size.

\begin{figure}[ht!]
\includegraphics[width=0.3\textwidth]{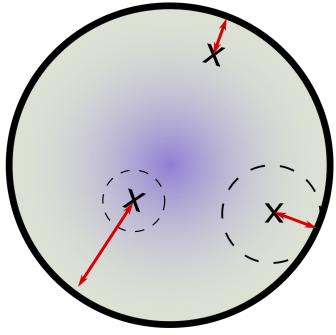}
\caption[]{The daughter grain consumes the HD-grains until it reaches a radius $r$ that makes it grow out of the (averaged) mother grain.}
\label{fig:nucgrowth}
\end{figure}

\paragraph{Grain growth} \label{par:gg} \mbox{}\\
For a growing/shrinking grain $k$, the (fractional) volume change due to interaction with a HEM during a time step $\Delta t$ is given by 
\begin{equation}
\Delta V^{HEM}_k= 4 \pi \phi^{\eta} r_k^2 v^{HEM}_k \Delta t.
\end{equation} 
Here $\phi^{\eta}$ is the fraction of the grain boundary that the grain $k$ shares with the HEM. For normal grains, there surface fractions are: $\phi^{LD}$ (between an LD-grain and the HD-HEM), $\phi^{HD}$ (between an HD-grain and the LD-HEM), $(1-\phi^{LD})$ or $(1-\phi^{HD})$ (between a grain and its own HEM) \cite{Mannheim2018}. The total volume change of a grain $\Delta V_k$ is then given by $\Delta V_k = \Delta V^{HD}_k + \Delta V^{LD}_k$. These equations are applied for all types of grain growth for the regular grains, however in case of shrinkage of a grain with respect to its own HEM, the corresponding volume change is determined such that volume is conserved within the HEM, i.e. $\sum_{i\in HD:\Delta V_{i}^{HD}>0} N_i \Delta  V_i^{HD}= - \sum_{i\in HD:\Delta V_{i}^{HD}<0}  N_i  \Delta V_i^{HD}$ for the HD-HEM \cite{Mannheim2018}. The interactions result in new grain sizes and new defect densities for all the regular grains. Next, the HD daughter grains grow at the cost of the HD-HEM. For these interactions a grain boundary surface fraction of 1 is used. This leads to a second change in grain size within the same time increment for the HD-grains. The HD daughter grains are not treated separately anymore from the moment that (1) their radius exceeds $\bar{r}_{HD}/4$ (see Figure~\ref{fig:nucgrowth}); or (2) once their bulk stored energy exceeds the threshold for becoming part of the HD-set themselves.

 For all recrystallization simulations, the grain boundaries between original grains are assumed to move more slowly due to pinning effects. For those boundaries, the pinned mobility constant $K_{m_0}$ is used instead of $K_m$, see also \cite{Mannheim2018}.

\paragraph{Nucleation}\mbox{}\\
The Gibbs free energy change over time for bulk nucleation is given by: 
\begin{align}
\frac{\partial{\Delta E}}{\partial{t}} & = \frac{\partial \Delta E }{\partial E^{B,HD}}\frac{d \Delta E^{B,HD}}{dt} + \frac{\partial \Delta E}{\partial r}\frac{dr}{dt}  =\nonumber \\
& = \frac{1}{K^B_a} \bigg[ - \frac{4 \pi}{3}r^3 \frac{d E^{B,HD}}{dt } - 4 \pi r^2 m (E^{B,HD}-E^B_0) (E^{HD} - E^B_0 )  \nonumber \\
	& + 6 \pi r m \gamma_b  (E^{B,HD} + \frac{4}{3} E^{HD} - \frac{7}{3} E^B_0 ) - 12 \pi m \gamma_b^2  \bigg] < 0,
\label{eq:dgb}
\intertext{whereas for necklace nucleation it is}
\frac{\partial{\Delta E}}{\partial{t}} & = \frac{\partial \Delta E }{\partial E^{B}}\frac{d \Delta E^{B}}{dt} + \frac{\partial \Delta E}{\partial r}\frac{dr}{dt} = \\
&  = \frac{1}{K^A_a} \bigg[ - \frac{4 \pi}{3}r^3 \frac{d E^B}{dt } - 4 \pi r^2 m (E^B-E^B_0) (E - E^B_0 ) + 6 \pi r m \gamma_b  (E^B + E - 2E^B_0 ) - 9 \pi m \gamma_b^2  \bigg] < 0.
\label{eq:dgn}
%& \frac{\partial \Delta E }{\partial E^{B}}\frac{d \Delta E^{B}}{dt} + \frac{\partial \Delta E}{\partial r}\frac{dr}{dt} = \nonumber \\
%& \frac{1}{K^A_a} \bigg[ - \frac{4 \pi}{3}r^3 \frac{d E^B}{dt } - 4 \pi r^2 m (E^B-E^B_0) (E - E^B_0 ) + 6 \pi r m \gamma_b  (E^B + E - 2E^B_0 ) - 9 \pi m \gamma_b^2  \bigg] < 0.
\end{align} The values for the derivatives $dE^{B,HD}/dt$ and $dE^B/dt$ are determined numerically. To obtain the radius of the nucleus $r_{nuc}$, the equation $\frac{\partial \Delta E}{\partial t}(r_0)=0$ is solved. The nucleated grain size $r_{nuc}$ is taken as 1.01 $r_0$, to ensure $v^{nuc}_{GB}>0$. If multiple solutions exist, then the largest value is adopted as the nucleus radius. The energy barrier that needs to be overcome for the formation of a stable grain is given by $E_{act} = \Delta E (r^{*})$, where $r^*$ is the solution to the static case $d \Delta E/dr = 0$. If there is no solution, then the nucleation rate is zero for that time step.\\
\\
Nucleation of new grains leads to small reductions of the sizes of other grains. The nucleation volume $V_{nuc}$ for grains that form in the grain interior is extracted from the HD-grains (all HD grain radii shrink with the same fraction). For necklace nucleation, the volume for nucleation is extracted similarly, but from both HEMs, where the share of each HEM is proportional to the surface fraction that it shares with the LD-grains (further detailed in Appendix~\ref{sec:appimp}). 

\paragraph{HEM allocation}\mbox{}\\
The original grain set (consisting of 50 original representative grains) is placed in the HD-HEM at the beginning of the simulation. All nucleated grains are initially defect-free and are thus placed in the LD-HEM. As soon as the energy density threshold $E_T$ is reached, they are placed in the HD-HEM. The maximum amount of nucleated representative grains in the HD-HEM is restricted to 50 to limit the computation time. Furthermore up to 50 nucleated representative grains may reside in the LD-HEM. In the case of bulk nucleation, maximum 16 out of the 50 representative grains in the LD-HEM may contain HD-daughter grains (which have low defect densities).\\
\\
When nucleation occurs, or when a representative grain is placed in the HD-HEM, it can be necessary to merge two representative grains together (two existing ones or an existing one with the nucleated one). Naturally, HD daughter grains and regular grains can only merge with grains of their own type. The two representative grains within a HEM that are most similar (based on the least-square difference of their bulk and surface stored energy densities $E^B$ and $E^S$ and based on the type of grain growth) are selected for merging. After merging, their properties in grain size and defect densities are averaged.

\paragraph{Numerical method} To solve the set of equations of the cluster dynamics model, Matlab's solver ode15s is used for each global time increment. Non-negativity of the solution is enforced with ode15s, and a relative tolerance of $10^{-3}$ is used.\\
\\
In Appendix~\ref{sec:appimp}, the solution procedure is set out in more detail, including a detailed description of the procedure with subincrements for handling grains that are completely consumed by their surroundings.

\subsection{Structure-property relations}
The yield strength is often considered as a key parameter for the divertor lifetime, as it gives an indication for the material hardening and embrittlement. Here, the evolution of the hardness is assessed, based on the predicted evolution of the microstructure.\\
\\
The hardening (which is proportional to the yield strength increase) of neutron-irradiated tungsten has been measured using Vickers micro-indentation tests for various doses and temperatures, as summarized in \cite{Fukuda2016,Hu2016}. TEM and positron annihilation spectroscopy (PAS) were used to link the hardness increase to the microstructural changes.  Hu \textit{et al.} \cite{Hu2016b} performed isochronal annealing testing following neutron irradiation of tungsten samples with neutrons at a temperature of 90 \textdegree C to 0.006 dpa and 0.03 dpa. They investigated the evolution of the Vickers hardness, vacancy clusters (with sizes ranging from monovacancies to several nm using PAS and TEM) and interstitial clusters (of at least 1 nm) as the temperature was increased. Their results suggest that vacancy clusters with sizes above 1 nm dominate the hardening, but small interstitial clusters were not taken into account in their analysis, as they are not detectable by PAS or TEM techniques. In \cite{Hwang2016}, ion-irradiation was used to study the hardening of pure W and of W-3\%Re at temperatures of 500 and 800 \textdegree C and at high doses of up to 5 dpa. Nano-indentation tests showed a saturation of the hardening at 1 dpa. For W-3\%Re, less hardening was found and this was attributed to the lower amount of voids (determined using TEM) \cite{Hwang2016}. \\
\\
In \cite{Hu2016}, Hu \textit{et al.} predicted the hardness increase based on the TEM-observed sizes and densities of the voids, dislocation loops and precipitates, by using the dispersed barrier hardening (DBH) model, using experimental results that were obtained in three different reactors HFIR, JMTR and JOYO \cite{He2006, Tanno2007, Fukuda2012, Hasegawa2013,Hu2016}. The defect barrier strengths, different for every defect type and size, were chosen such that the DBH-model and the measured Vickers hardness adequately matched for most samples. Their results suggest that the hardness increase is dose dependent but not strongly temperature dependent. Voids of over 4 nm were identified as the main contributors to hardening in most cases, except for samples that received doses of above 1.5 dpa in the HFIR-reactor. In the latter case, the precipitates were the dominant defect contributing to hardening, according to their DBH-model \cite{Hu2016}. Huang \textit{et al.} \cite{Huang2018} simulated the hardness increase in the reactors DEMO, HFIR and JOYO using stochastic cluster dynamics (SCD), including Re-transmutations, for temperatures of 400-800 \textdegree C and for doses up to 2 dpa. They found precipitates to be the main contributor to hardening, already for doses of 0.5 dpa, for each of the three reactors. Between voids and dislocation loops, the latter contributed more to the hardening than the voids, in most of the cases. The predicted hardness increases were in the order of 5-20 GPa. Without explicit formation of precipitates, the measured displacement-induced irradiation hardening does generally not exceed 4 GPa (see Hu \textit{et al.} \cite{Hu2016}). In \cite{Marian2012}, stochastic cluster dynamics (SCD) was used to predict hardening in the fission reactor JOYO and in ITER. In this case, the production of He gas was taken into account but not the Re-precipitation. Low hardening values were found in the order of 100 MPa.\\
\\
The yield strength depends on the concentrations of the vacancy and interstitial clusters, the dislocation network density and the grain size (grain boundary density). Each of these defect types form obstacles that restrict dislocations in their motion. The yield strength is then given by:
\begin{align}
\sigma_y 	&  = \sigma_0 + \Delta \sigma (r) + \Delta \sigma (\rho,C_j),
\intertext{with $\sigma_0$ the yield strength of coarse grained recrystallized tungsten, $\Delta \sigma(r)$ the grain boundary hardening contribution and $ \Delta \sigma (\rho,C_j)$ the hardening contributions of the dislocation network density and the vacancy and interstitial clusters. The Hall-Petch relation describes the grain size dependence, the Taylor relation \cite{Taylor1934, Davoudi2014} the effect of dislocation hardening and the Dispersed Barrier Hardening Model \cite{Hu2016} the contributions of the vacancy and interstitial clusters. Combining the contributions in a similar manner as in \cite{Garner1981,Hansen2004} leads to:}
\sigma_y 	&  = \sigma_0 + \frac{k_1}{\sqrt{r_{g}}} + M\alpha_T \mu b \sqrt{\rho} +  \sqrt{\sum\limits_{j} \Delta \sigma_j^2},
\end{align}with $\Delta \sigma_j = M\alpha_j \mu b \sqrt{2C_jr_j}$. In the grain size contribution, $r_{g}$ is the grain radius, $r_j$ is the radius of the defect type $j$ that forms the barrier and $k_1$ is a constant. $k_1$ should be dependent on the temperature, but here, a value of 0.099 MPa m$^{1/2}$ is used, based on experimental results by Vashi \cite{Vashi1970}. For $\sigma_0$, 64 MPa is used (the value for recrystallized tungsten at 1200 \textdegree C, taken from \cite{IMPH2013}). In the contributions of the various defect types, $M$ is the Taylor factor, $\mu$ is the shear modulus, $\alpha_j$ is the defect barrier strength factor (for perfectly strong barriers to dislocations, $\alpha = 1$) and $C_j$ is the defect concentration. The parameter values are given in Table~\ref{tab:parval} (Appendix~\ref{app:par}) and in section~\ref{sec:lu}. In the simulations, the parameters $\mu$ and $b$ are treated as temperature independent and also given in Table~\ref{tab:parval}.  The yield strength increase is directly related to the Vickers hardness increase \cite{Hu2016}. In this work, only the qualitative aspects of the hardness evolution under the combined effects of irradiation, recrystallization and grain growth are studied. Therefore, a hardness indicator $\mathcal{I}_H$ is introduced:
\begin{align}
\mathcal{I}_H = \frac{\sqrt{\rho} + \frac{M}{\alpha_T}\sqrt{\sum_j2\alpha^2_j C_j r_j}}{\sqrt{\rho_0}},
\end{align}
where $\rho_0$ is the initial network dislocation density. The Hall-Petch effect is neglected in this expression. 

%Because of the simplified nature of the multi-scale model of the current work, it is chosen to study the evolution of hardness indicator $\mathcal{I}_H$, which is defined as:

%%%%%%%%%%%%%%%%%%%% 

\section{Results}
This section analyses how the microstructural evolution of tungsten, under combined heat and neutron loads, is affected by loop incorporation and bulk nucleation, as an extension to the work done in \cite{Mannheim2018}. In section~\ref{sec:sg}, the effects of loop incorporation on the defect concentration distribution and on the defect energy evolution of individual grains are studied. Next, in section~\ref{sec:lu}, the temperature-dependent neutron-induced recrystallization with loop incorporation is explored and a prediction of the irradiation hardening during the process is made. The effects of bulk nucleation are described in section~\ref{sec:bn}.

\newpage

\subsection{Single grain} \label{sec:sg}
\begin{figure}[H]
\subfloat[]{\includegraphics[height=0.21\textheight]{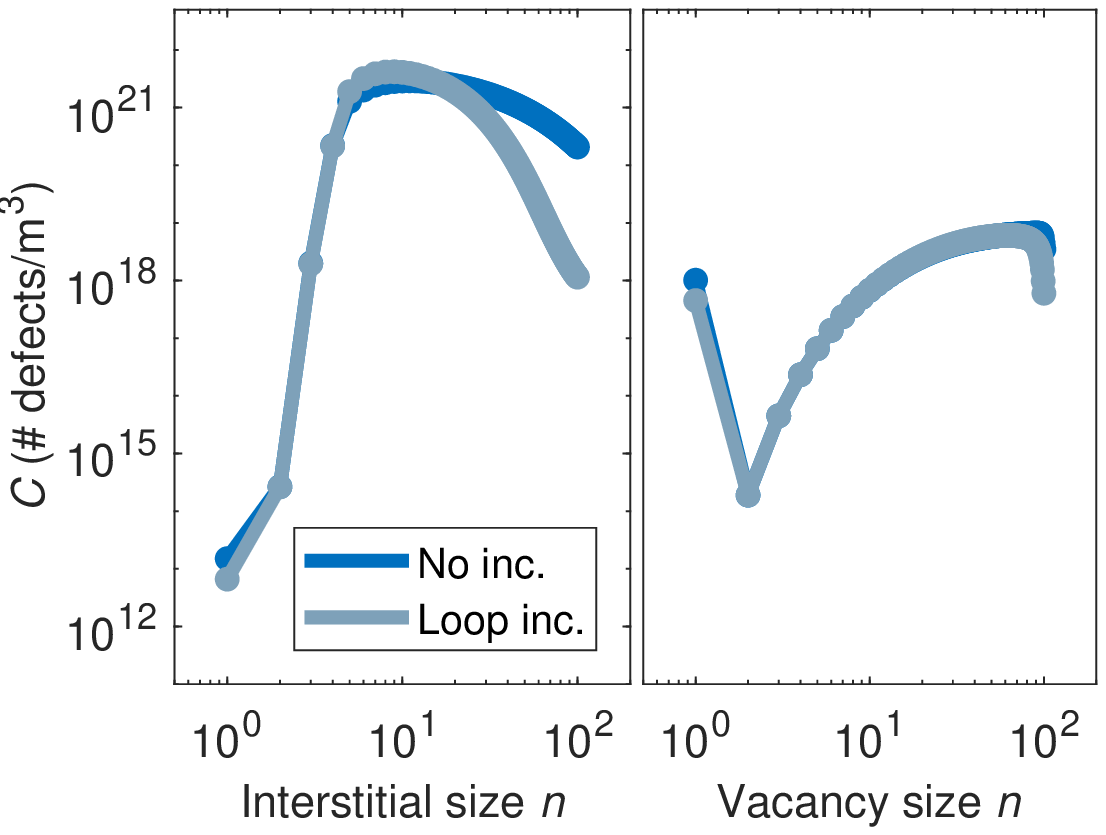}} \hspace{0.2 cm}
\subfloat[]{\includegraphics[height=0.227\textheight]{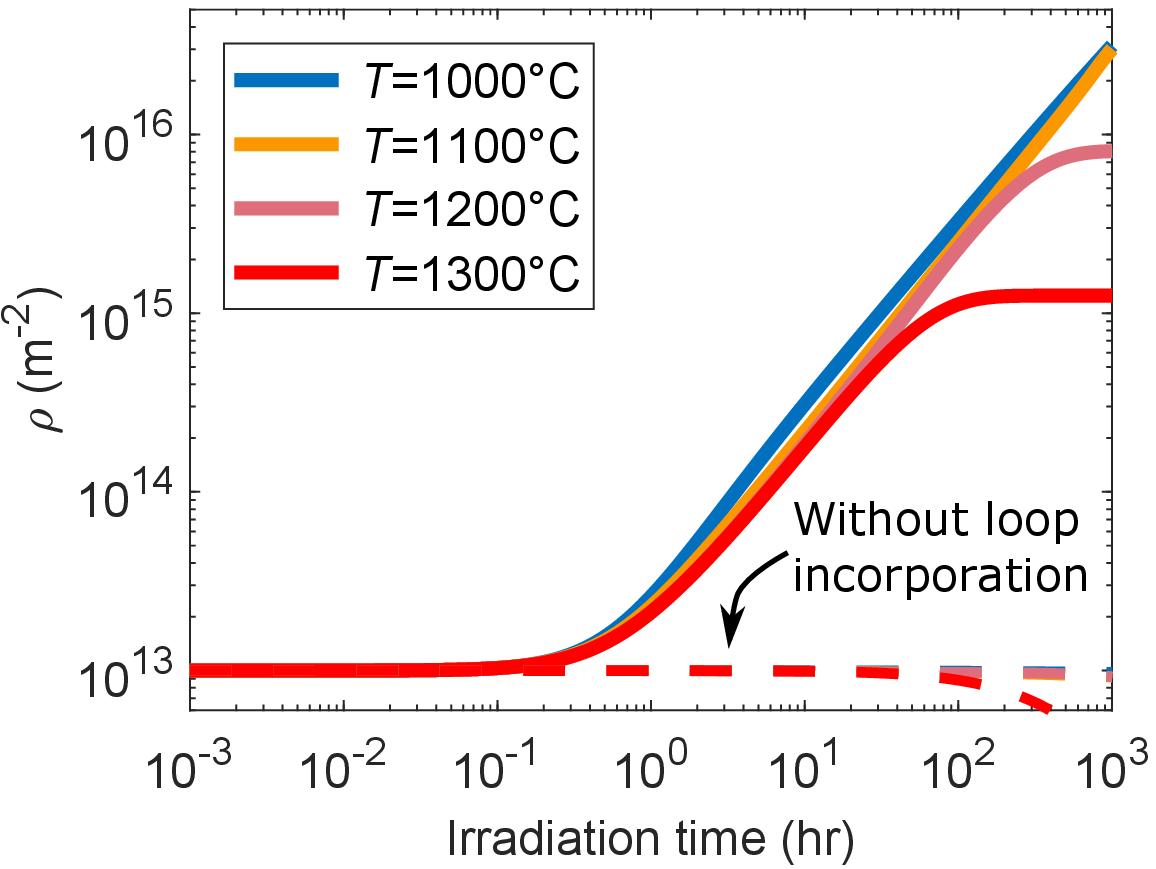}}
%\subfloat[]{\includegraphics[height=0.22\textheight]{LoopImproved_Tdep_Dislocationdensity.eps}}
\caption[]{Simulation of displacement damage in an irradiated grain of $r$ = 3 $\mu$m: (a) the concentration distribution of the interstitial and vacancy defect clusters after irradiation for 100 hours at a temperature of 1100 \textdegree C; (b) the evolution of the dislocation network density at several temperatures, as a consequence of Bardeen-Herring sources, dipole annihilation and incorporation of prismatic dislocation loops into the dislocation network.}
\label{fig:cdist}
\end{figure}

First, the effect of the incorporation of prismatic loops on the microstructural evolution of tungsten at high irradiation temperatures is investigated using the cluster dynamics model only. In Figure~\ref{fig:cdist}a, the concentration distribution in a grain of 3 $\mu$m after 100 hrs of irradiation at a temperature of 1100 \textdegree C is shown, with and without loop incorporation, using an initial dislocation density of $\rho=10^{13}$ m$^{-2}$.
%\footnote{It was verified that, compared to $N_{max}$=1000, the evolution of the atomic defect fraction $f$ was in reasonable agreement for most damage rates (e.g. varying $f_{rem}$ from 0.001 to 1) at each of the irradiation temperatures (1000-1300\textdegree C).}.

\begin{figure}[ht!]
\centering
\subfloat[]{\includegraphics[height=0.22\textheight]{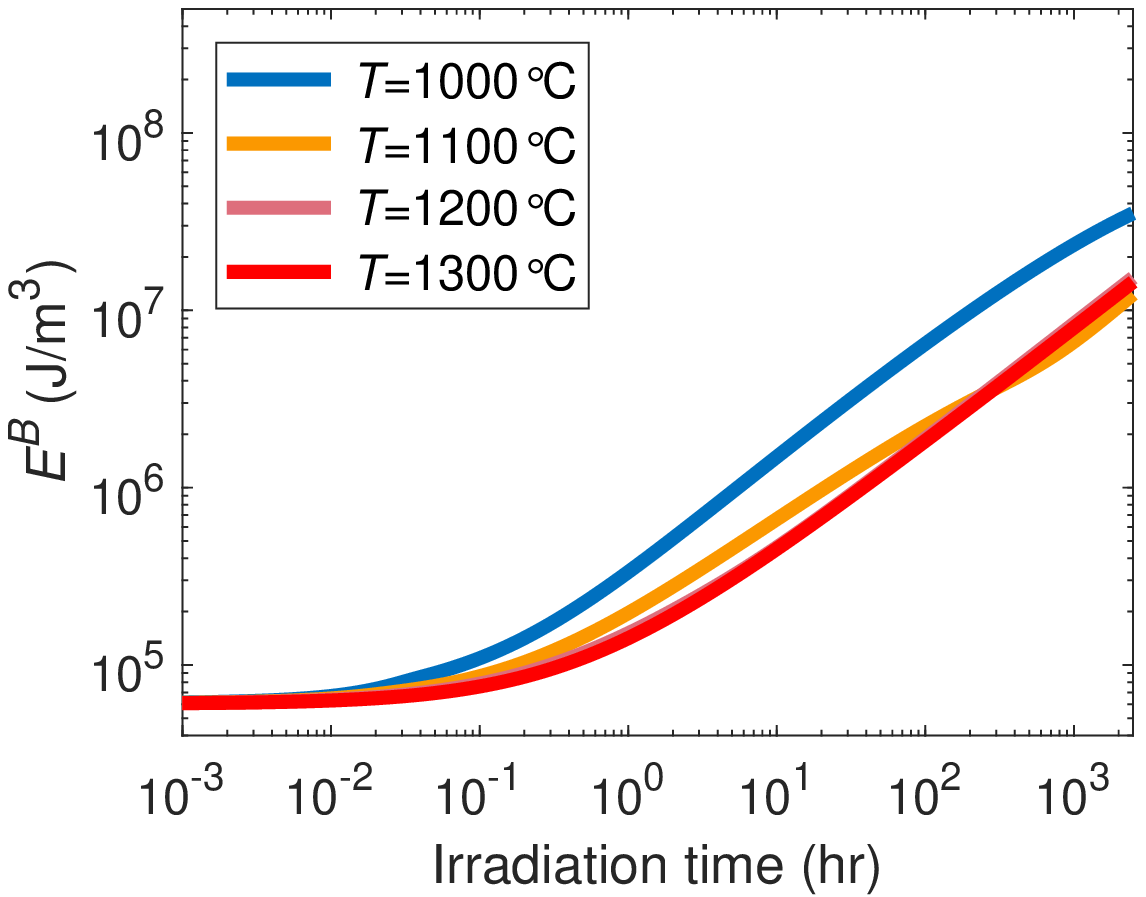}} \hspace{0.2 cm}
\subfloat[]{\includegraphics[height=0.22\textheight]{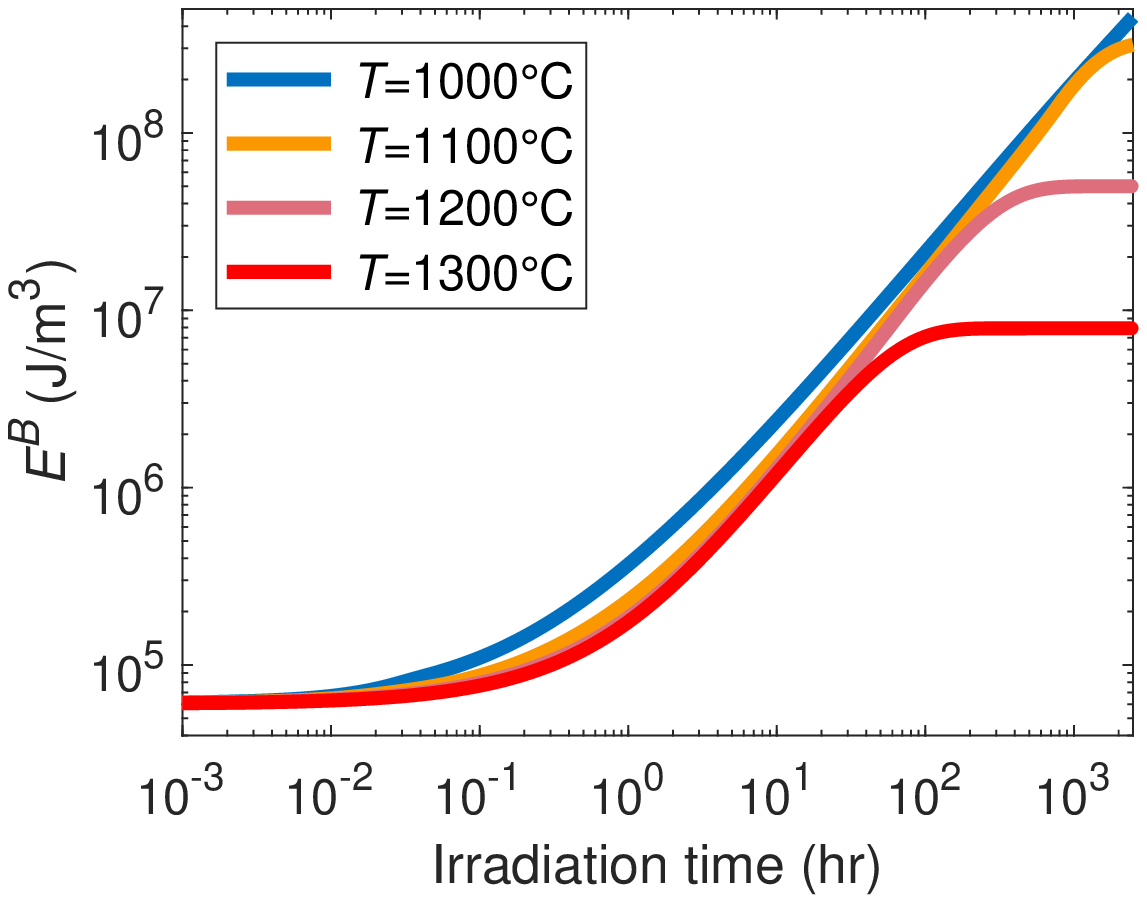}}
\caption[]{Evolution of the bulk stored energy density $E^B$ in a single grain with a radius of 3 $\mu$m at several irradiation temperatures: (a) without and (b) with loop incorporation.}
\label{fig:sg}
\end{figure}

As visible in Figure~\ref{fig:cdist}a, loop incorporation leads to a decrease in the concentration of the largest interstitial clusters and to an increase of the dislocation network density (see Figure~\ref{fig:cdist}b). Note that without the loop incorporation mechanism, the dislocation density in the simulations stays close to $10^{13}$ m$^{-2}$. The strength of the sinks for the mobile defects (the grain boundaries and the dislocations) thus increases and with that, the concentrations of the single vacancies and single SIAs decrease. The concentration of the interstitial clusters of intermediate size ($n=5$ to 25) increases slightly, whereas the concentration of the largest vacancy clusters decreases somewhat. The dislocation density, shown in Figure~\ref{fig:cdist}b, rises considerably for all simulated temperatures when loop incorporation is included, approaching the physical limit for the dislocation density in metals, $10^{17}$ m$^{-2}$ \cite{Cotterill1977}. \\
\\
Figure~\ref{fig:sg} shows the evolution of the bulk stored energy density inside the grain, with and without the loop incorporation mechanism.
Figures~\ref{fig:sg}a and~\ref{fig:sg}b reveal that at longer irradiation times, the temperature dependence of the bulk stored energy density $E^B$ becomes more pronounced when loop incorporation is included in the simulations, reaching much higher values for the lower temperatures (1000 \textdegree C - 1200 \textdegree C) and a similar/lower value at 1300 \textdegree C. Comparison of Figure~\ref{fig:cdist}b and Figure~\ref{fig:sg}b indicates that the value of $E^B$ is largely determined by the dislocation network density. Furthermore, when loop incorporation is included, saturation of the stored energy density in the single grain occurs. For e.g. 1300 \textdegree C, this occurs within 100 hours and for 1100 \textdegree C, saturation is reached at 2500 hours. \\
\\
As an example, the influence of the damage rate and  $N_{max}$ on the results is shown for $T=1000$ \textdegree C in Figure~\ref{fig:damrate} in Appendix~\ref{app:damrate}.
%It was chosen to use a maximum cluster size $N_{max}$ of 100 for all the simulations in this paper, whereby it has been verified that a larger $N_{max}$ leads to similar defect concentration distributions and stored defect energies. 
%Without loop \textcolor{blue}{incorporation}, a much larger value for $N_{max}$ would be required in order to achieve convergence in terms of $E^B$.
\\
\\
%\textcolor{red}{In order to select $N_{max}$, both the damage rate and the maximum cluster size were varied for each of the irradiation temperatures 1000-1400 \textdegree C. Simulations were run for $N_{max}=100$ and $N_{max}=1000$ and using various damage rates, on the order of $10^{-12} - 10^{-8}$ point defect pairs/s. The evolution of the defect fraction as a function of the amount of applied damage was compared for each of the simulation conditions. For temperatures of $1000-1300$ \textdegree C, it was concluded that $N_{max}$=100 gives a similar evolution of the defect fraction differed (no more than a factor 2 difference for each of the damage rates). At an irradiation temperature of 1400 \textdegree C, at slower damage rates, $N_{max}=100$ and 1000 led to strongly different results. It was chosen to omit this irradiation temperature from the calculations. Furthermore, it was chosen to use $N_{max}=100$ for all the other temperatures.}

%%%%%%%%%%%%%%%%%%%%%%%%%
\subsection{NIRX with loop incorporation} \label{sec:lu}
\begin{figure}[H]
\subfloat[]{\includegraphics[height=0.22\textheight]{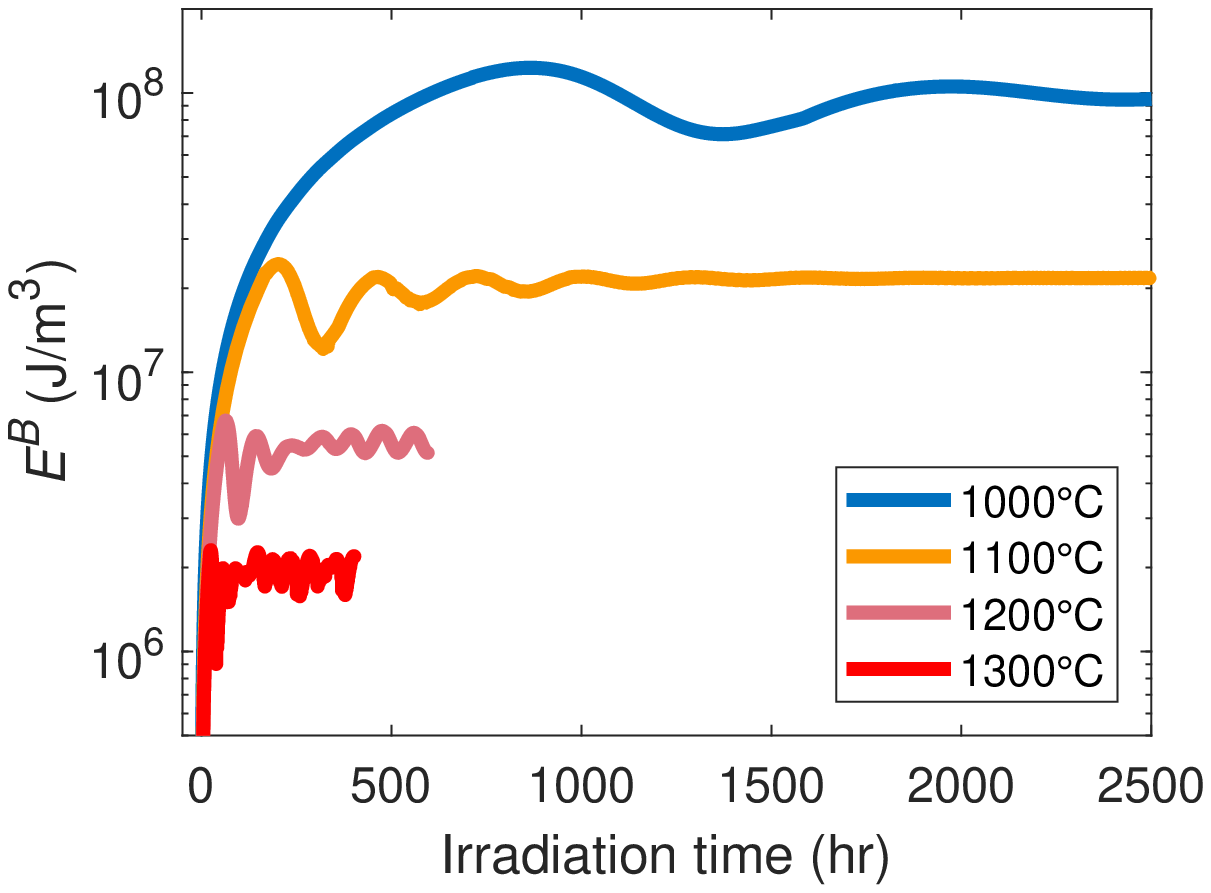}} \hspace{0.1cm}
\subfloat[]{\includegraphics[height=0.22\textheight]{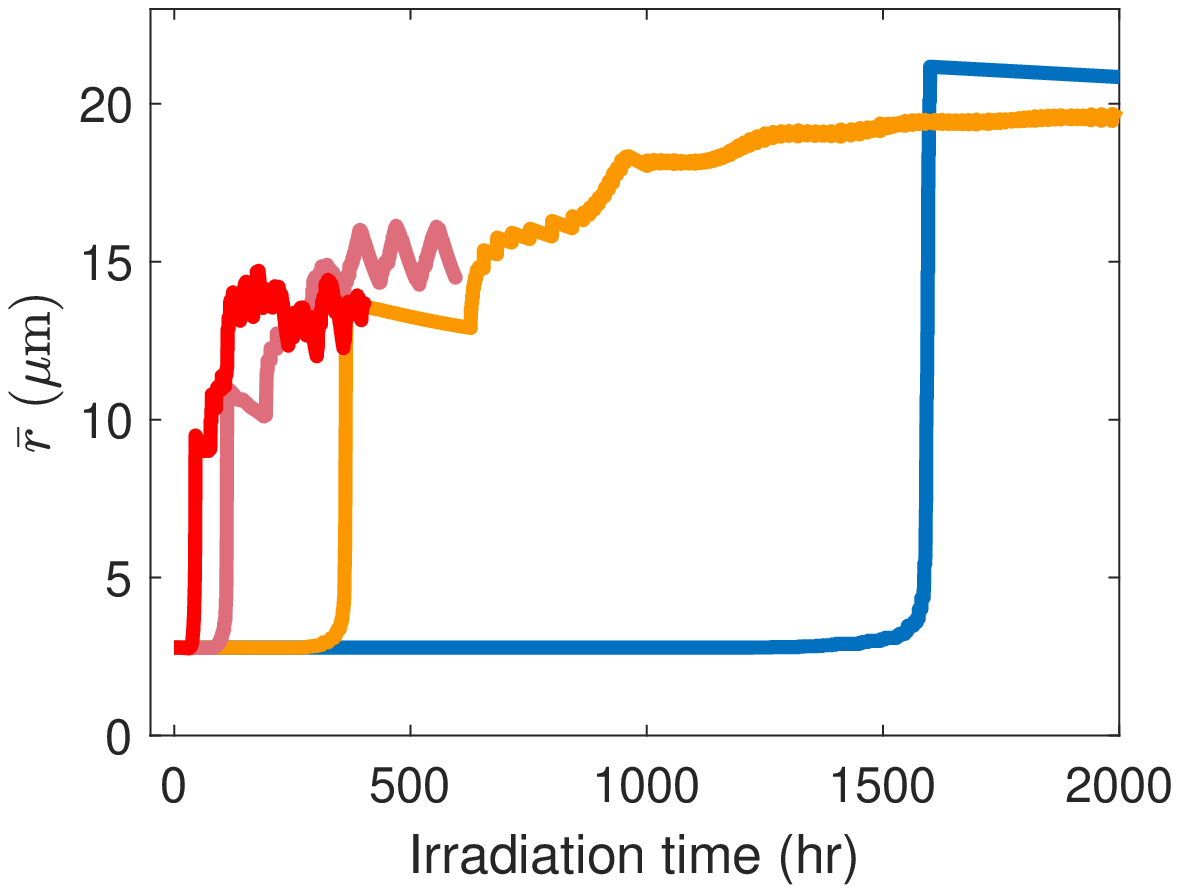}} 
\caption[]{Evolution of (a) the bulk stored energy density $E^B$ and (b) the average grain size $\bar{r}$ during neutron-induced recrystallization simulations.}
\label{fig:loopunf}
\end{figure}

\begin{figure}[ht!]
\subfloat[]{\includegraphics[height=0.22\textheight]{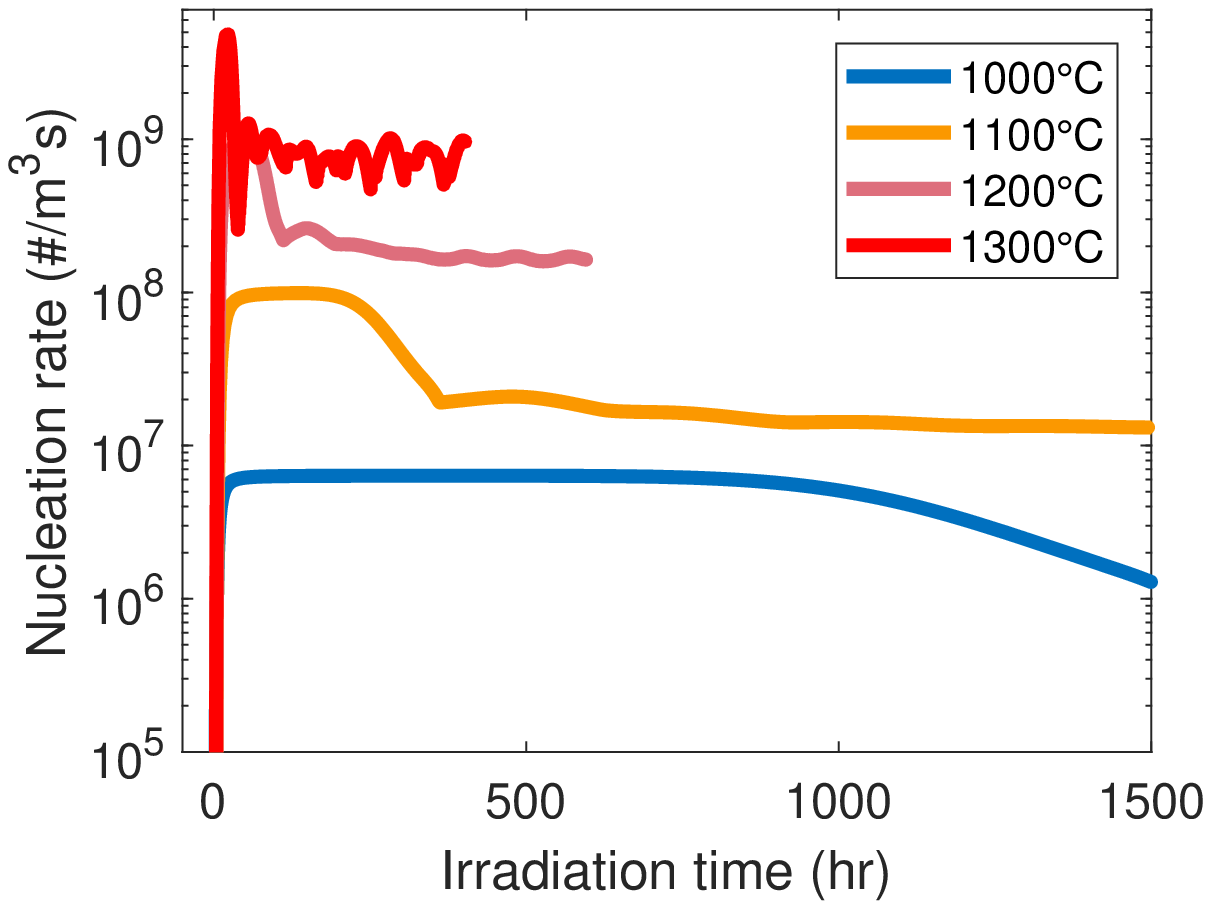}} \hspace{0.05cm}
\subfloat[]{\includegraphics[height=0.22\textheight]{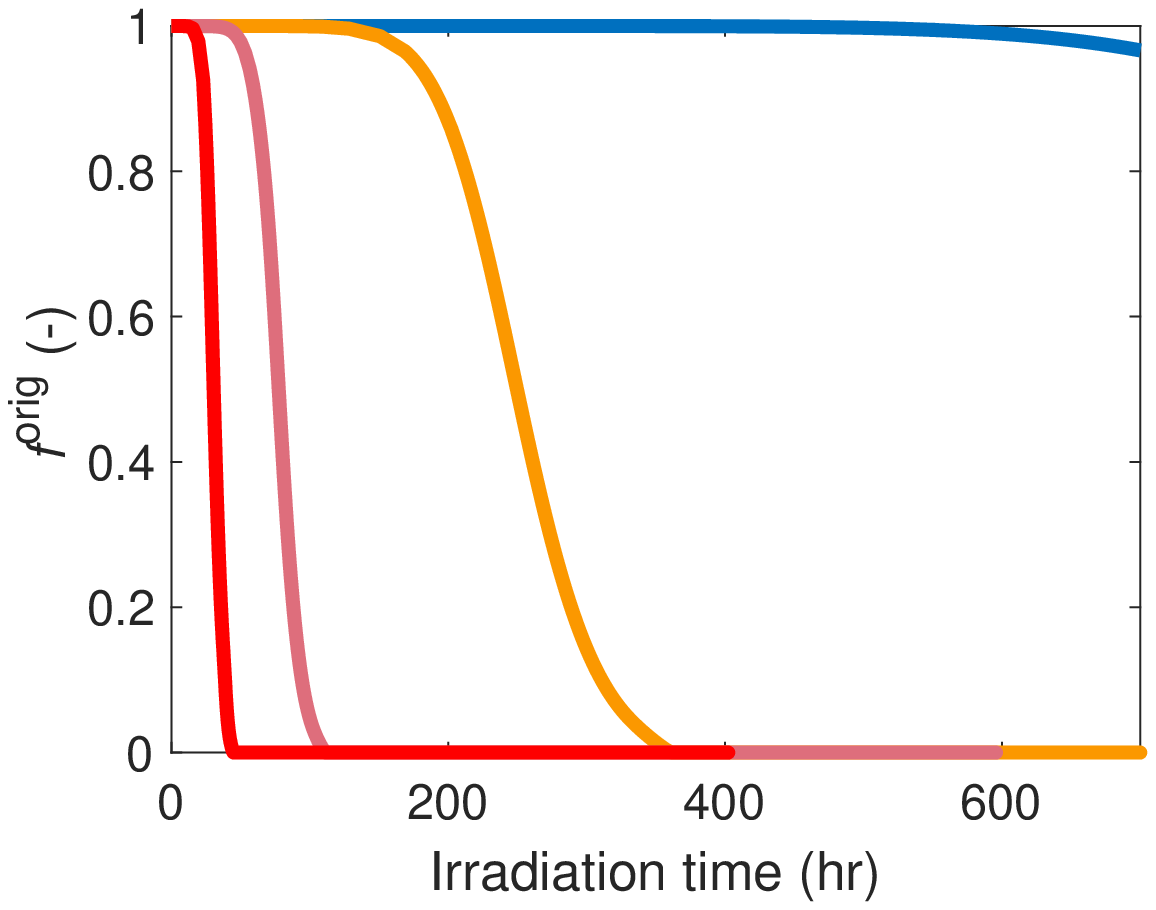}} \vspace{0.01cm}
\subfloat[]{\includegraphics[height=0.22\textheight]{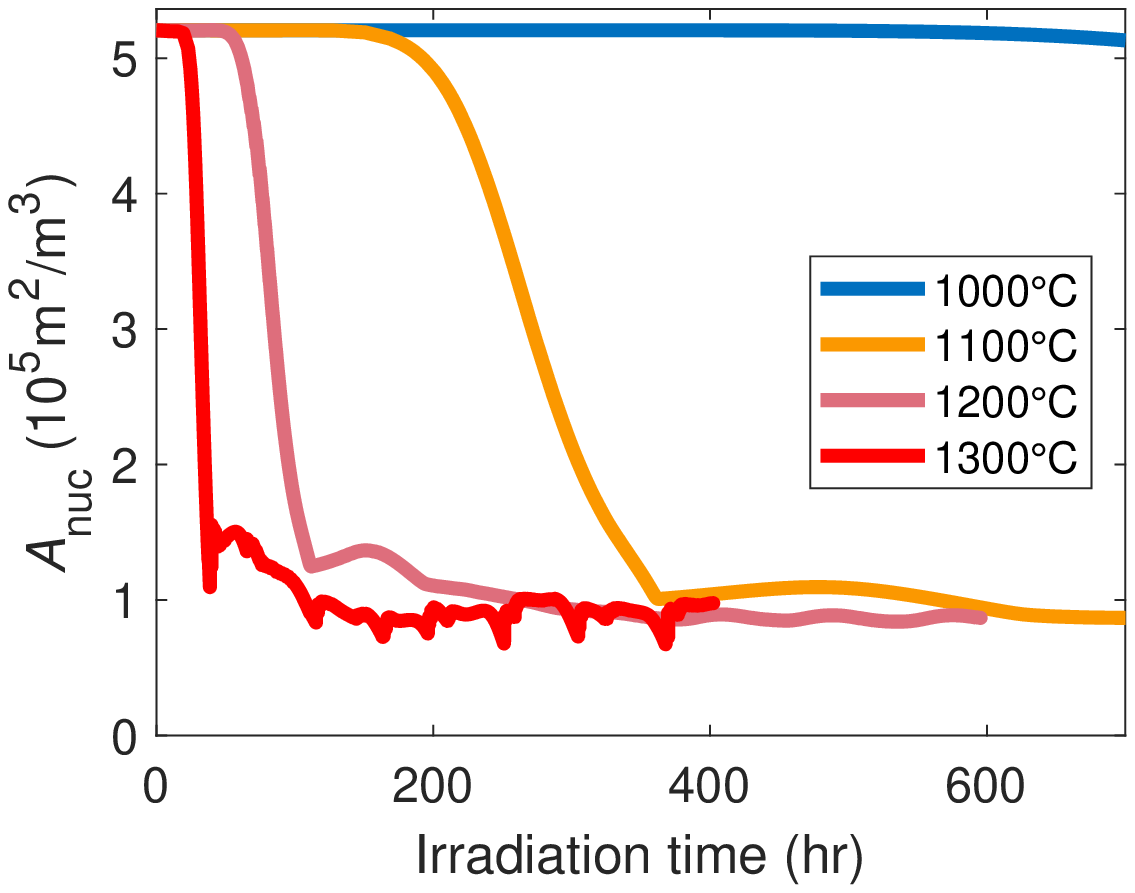}}
\caption[]{Evolution of the nucleation rate (a), of the fraction of the volume that corresponds to the HD-medium (b) and of the surface area per unit volume that is available for nucleation (c).}
\label{fig:loopnuc}
\end{figure}

Previously, the competition between damage and recovery during the microstructural evolution of tungsten under heat and neutron loads was already investigated in \cite{Mannheim2018}, omitting loop incorporation. It was found that the original grains, which accumulate defect energy during irradiation, all shrink simultaneously at the cost of newly nucleated grains. This simultaneous loss of the original grains triggered temporal fluctuations in the average stored energy density in the microstructure $E^B$. Now that loop incorporation is included, $E^B$ shows fluctuations that are clearly periodic, see Figure~\ref{fig:loopunf}a. Before, only the first undulation was clear. The undulations in $E^B$ correspond to the cyclic recrystallization of the entire irradiated microstructure \cite{Mannheim2018}. For the simulations displayed in Figure~\ref{fig:loopunf}, an energy density threshold $E_T=10^6$ J/m$^3$ was used. In all original grains the initial dislocation density was $\rho=10^{13}$ m$^{-2}$. The defect-free volume has a dislocation density $\rho^{eq}=10^9$ m$^{-2}$. For the initial vacancy and interstitial defect concentrations, thermal equilibrium values were used. The 50 original representative grains had a number average radius of $\bar{r}$ = 3 $\mu$m and a standard deviation of $\sigma_r$ = 0.5 $\mu$m.
The time period characterizing cyclic recrystallization is temperature-dependent. Over time, the behaviour of the individual grains becomes less synchronized, and the amplitude of the fluctuations of $E^B$ decreases and the stored energy density evolves to a temperature-dependent equilibrium value. For temperatures of 1200 \textdegree C - 1300 \textdegree C, $E^B$ oscillates around the equilibrium value.\\
\\
The average grain size, shown in Figure~\ref{fig:loopunf}b, does not reveal a strong temperature dependence. Both the evolution of the grain size and of the defect density depend strongly on the nucleation rate, see Figure~\ref{fig:loopnuc}a. 
At each temperature the nucleation rate shows a clear drop when the nucleation surface area has decreased, which occurs when the grain size has increased (when the original microstructure has been replaced). The nucleation rate depends on the choice of the energy density threshold $E_T$, since all the grain boundaries of the HD-grains act as nucleation surfaces. From Figure~\ref{fig:loopnuc}b it becomes clear that the fraction $f^{HD}$ of the total microstructural volume that resides in the high defect density medium is close to 1 most of the time for the temperatures up to 1200 \textdegree C. From 1200 \textdegree C, the surface area that is available for nucleation per unit volume (Figure~\ref{fig:loopnuc}c) fluctuates, which leads to variations in the nucleation rate and in $E^B$.

\begin{figure}[H]
\subfloat[]{\includegraphics[height=0.22\textheight]{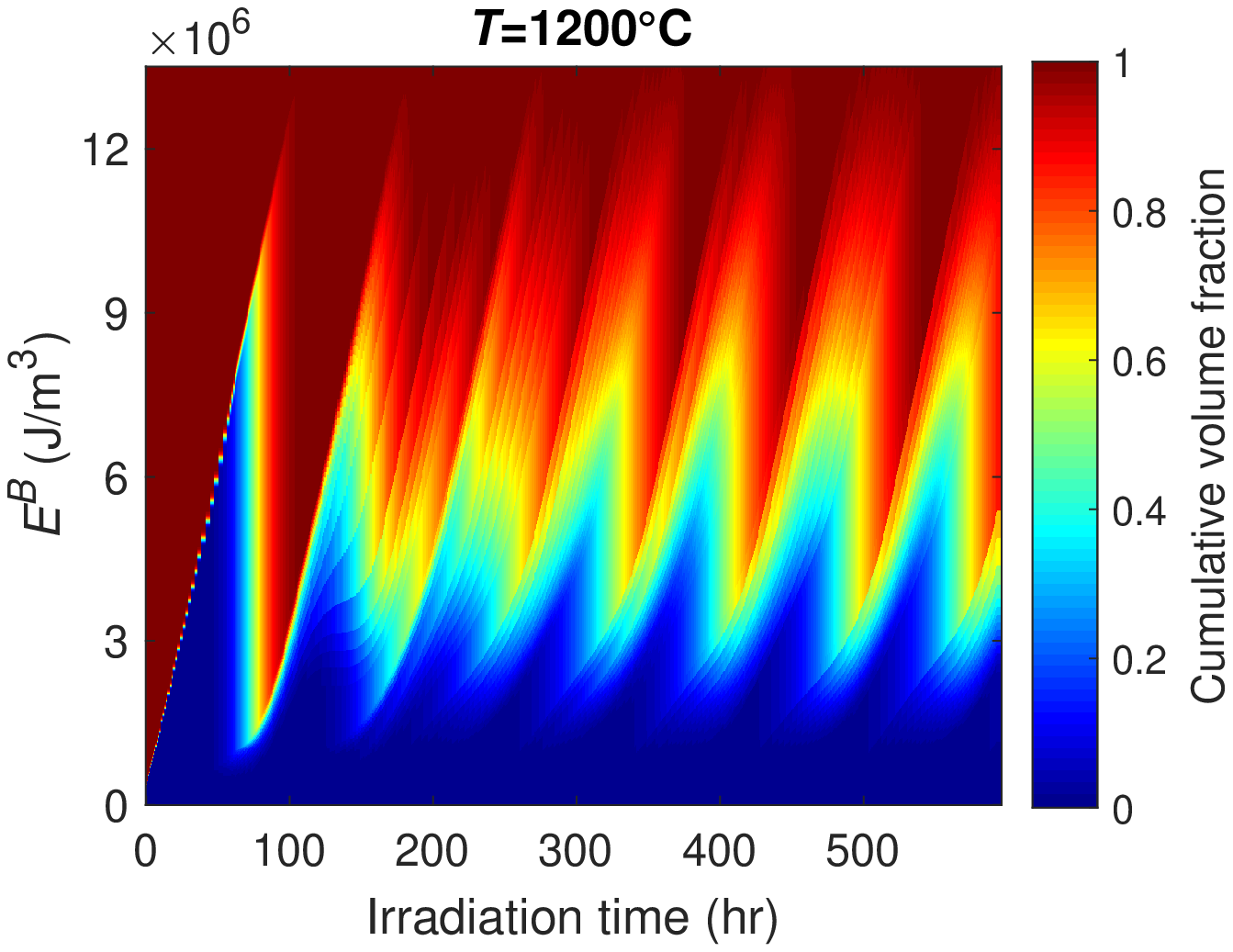}} \hspace{0.2cm}
\subfloat[]{\includegraphics[height=0.22\textheight]{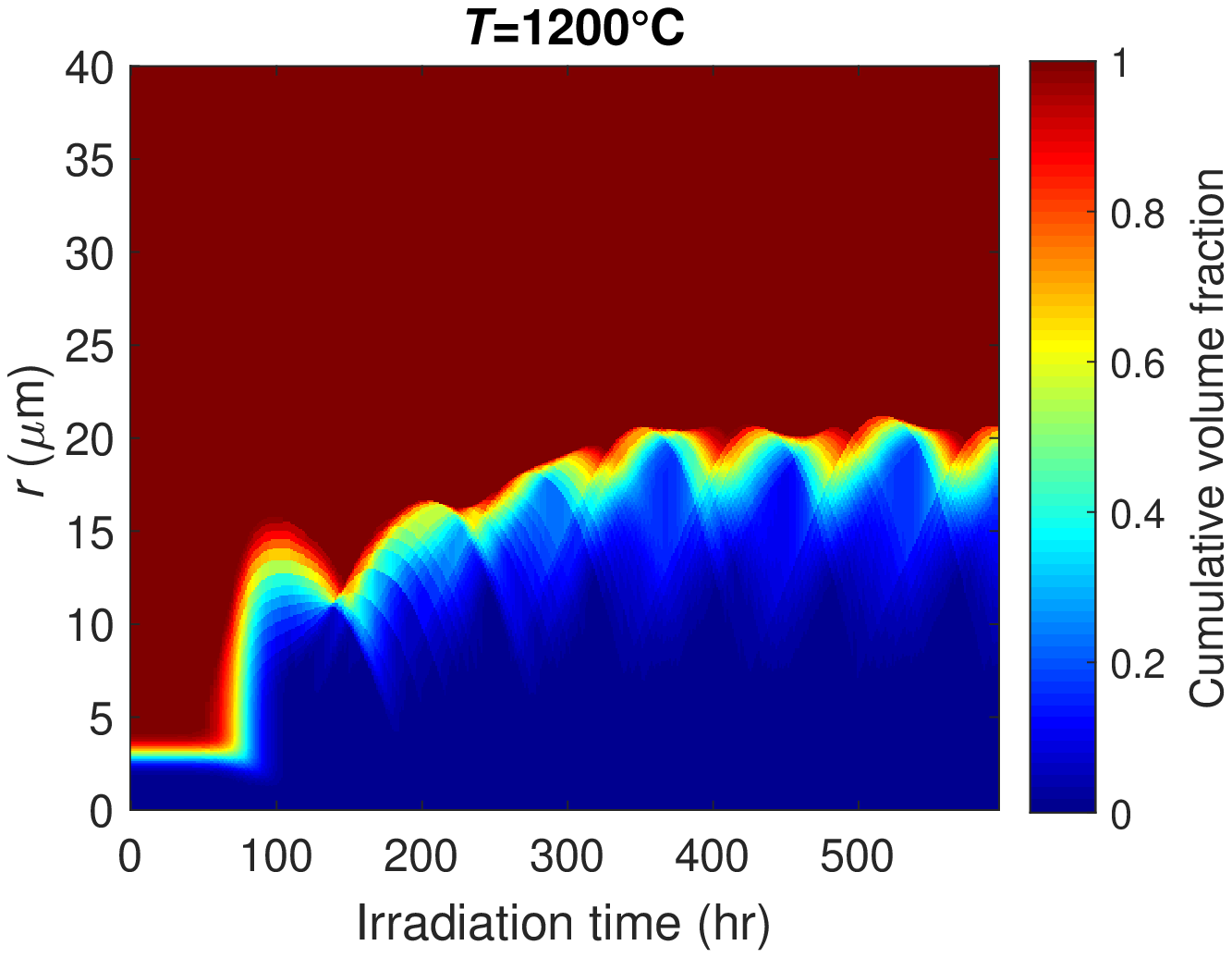}} 
\caption[]{Evolution of (a) the bulk energy density distribution and (b) the grain size distribution at a temperature of 1200 \textdegree C, in terms of cumulative volume fractions.}
\label{fig:erdist}
\end{figure}

\paragraph{Distributions in grain size and defect density} \mbox{}\\
In Figure~\ref{fig:erdist}, the evolution of the bulk energy density distribution and grain size distribution are shown for an irradiation temperature of 1200 \textdegree C.
Over time, the distributions broaden, i.e. the microstructure becomes more heterogeneous and recrystallization becomes more gradual. This explains the decrease in the amplitude of $E^B$ in Figure~\ref{fig:loopunf} after the first 200 hours. The evolution of the grain size and energy distributions at the other simulated temperatures are reported in Figure~\ref{fig:gdist} and Figure~\ref{fig:edist} in Appendix~\ref{app:cumdist}. 

\begin{figure}[H]
\subfloat[]{\includegraphics[height=0.23\textheight]{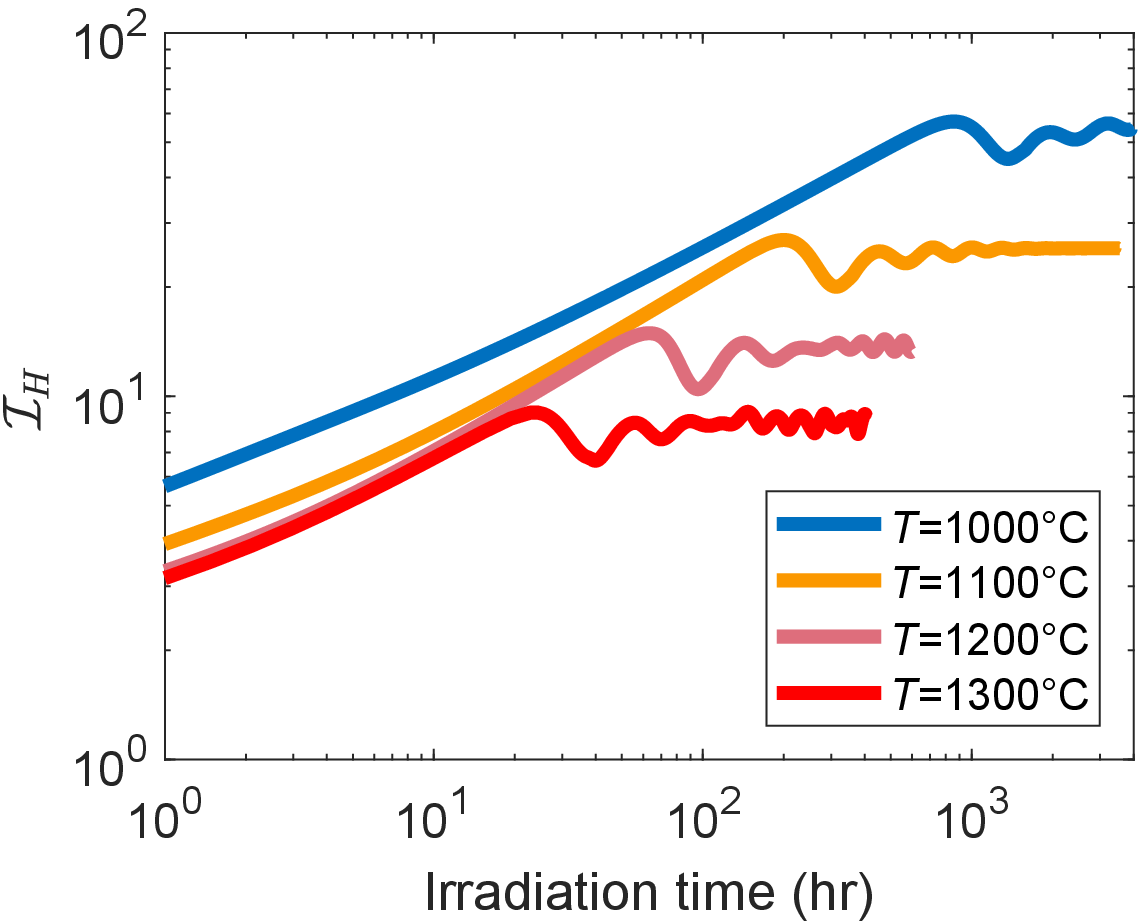}} 
\subfloat[]{\includegraphics[height=0.23\textheight]{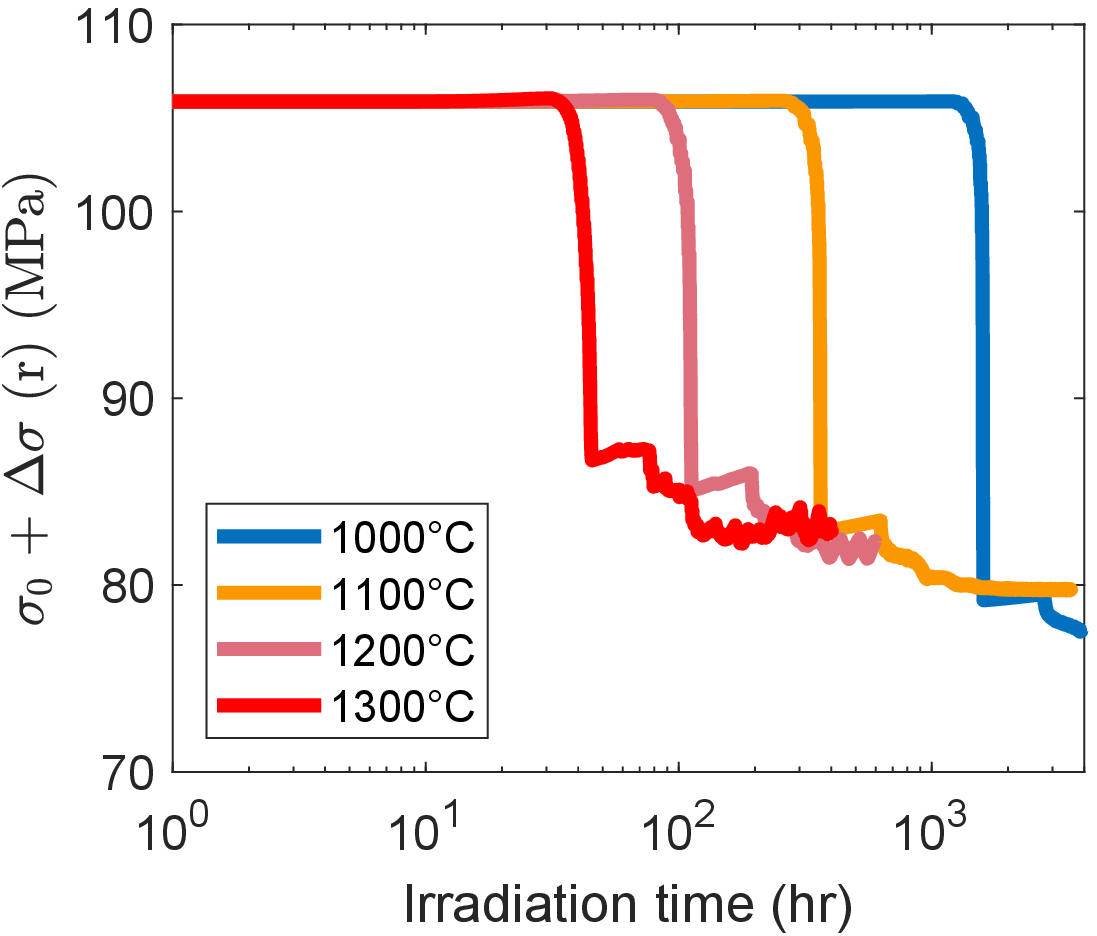}} \vspace{0.1cm}
\subfloat[]{\includegraphics[height=0.22\textheight]{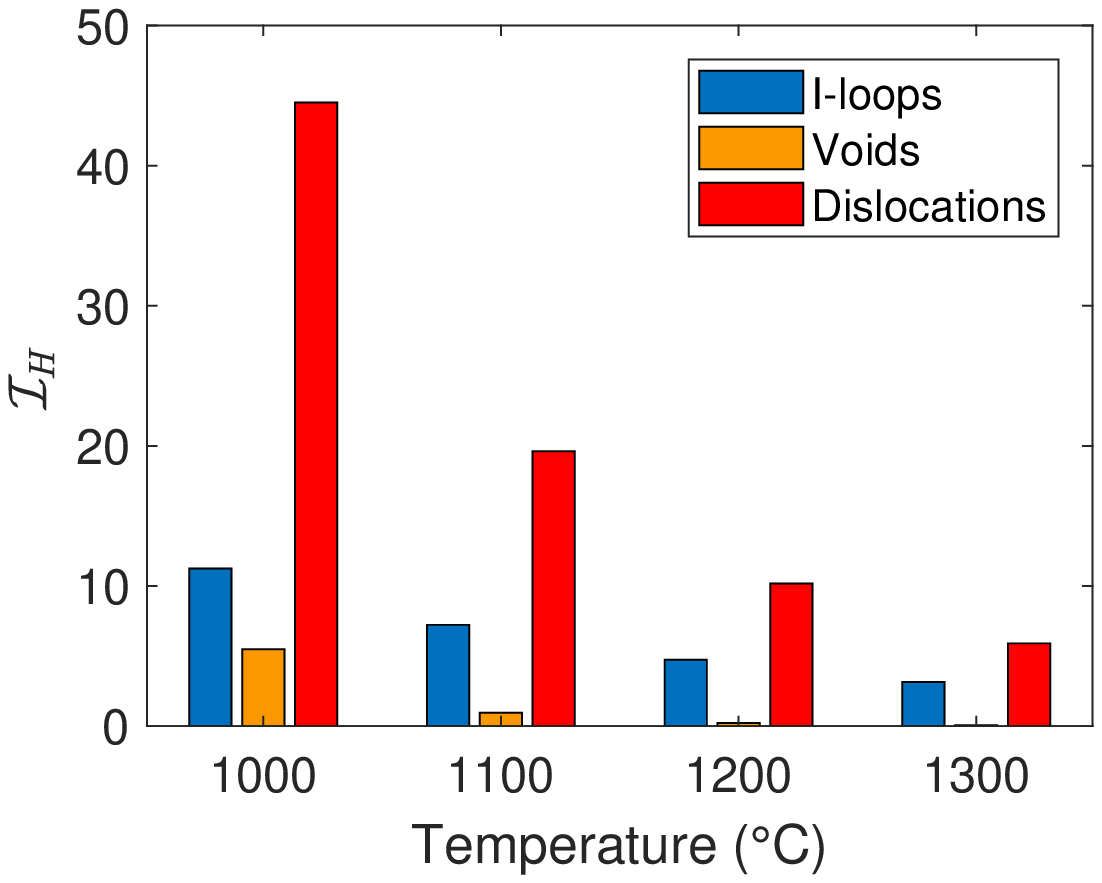}}
\caption[]{(a) Evolution of the irradiation hardening indicator as a function of applied damage; (b) Effect of the grain size on the hardening; (c) Contributions of the various defect types to thehardening indicator, taken at time instances where $\mathcal{I}_H$ reaches its maximum.}
\label{fig:hardhu}
\end{figure}

\paragraph{Irradiation hardening}\mbox{}\\
The evolution of the irradiation hardening was predicted, using the values in Table~\ref{tab:dbh} for the DBH-model for the indicated cluster sizes. In Figure~\ref{fig:hardhu}a, the results are shown, in comparison with results from Vicker's hardness tests on samples that were irradiated in the fission reactors HFIR, JOYO and JMTR \cite{Hu2016}. Figure~\ref{fig:hardhu}b exclusively shows the (negligible) effect of the grain size on the evolution of the yield strength, based on the Hall-Petch relation. The contributions of the different defect types to the irradiation hardening indicator, taken at time instances of the maximum value for the hardness indicator, are shown in Figure~\ref{fig:hardhu}c. 

\begin{table*}[ht!]
\caption[]{Barrier strengths as used in the DBH-model, based on \cite{Hu2016}.}
\begin{tabular}{l | c | c  | c }
 			& Barrier strength factor $\alpha$ 	&     Diameter (nm) 		&  Cluster size $N$ 		\\ \hline
Interstitial loop       & 	0.15 					& 	1.0 - 2.7	 	& 14 - 100 			\\
Void 			& 	0.25 					& 	1.0 - 1.4 		& 33 - 100 			
\label{tab:dbh}
\end{tabular}
\end{table*}

Under the selected neutron displacement damage rate, the hardness indicator increases during the first 20-900 hours of irradiation (for 1300\textdegree C and 1000\textdegree C). Subsequently, cyclic recrystallization leads to a periodic increase and decrease of the hardness indicator. The hardness indicator reaches a maximum value of 10-80 times the initial value (=1).\\
\\
Figure~\ref{fig:hardhu}c reveals it can be seen that the contribution of the dislocation network density to the hardening is the highest at each of the simulated temperatures, which is logical based on the high dislocation density reached, see Figure~\ref{fig:cdist}b. In \cite{Hu2016b}, it is suggested that the large voids are the main defects that are responsible for the hardness increase. Therefore, additional simulations were performed using the cluster dynamics model at 1100 \textdegree C on a single grain of 3 $\mu$m for increasing maximum vacancy cluster size $N_{max}$. The results (not displayed here) showed that the evolution of the hardness is almost converged for $N_{max}=100$, as expected, since the same holds for $E^B$, indicating that large clusters would not play a significant role in the hardness prediction for this model. \\
\\
In the present analysis, the influence of the grain boundaries on the yield strength is taken into account only partly. In \cite{Was2007}, it is pointed out that at high temperatures, the grain boundary loses strength relative to the grain interior. Furthermore, recrystallization and grain growth may lead to an increase of impurities at the grain boundaries, some of which may cause embrittlement at the grain boundaries \cite{Setyawan2012,Scheiber2016}. 

%%%%%%%%%%%%%%%%%%%%%%%%%
\subsection{NIRX with bulk nucleation} \label{sec:bn}
The effect of nucleation in the bulk (on top of necklace-type nucleation) on the microstructural evolution is next studied for an irradiation temperature of 1100 \textdegree C. The parameter $K^B_a$ for the reduction of the activation energy for nucleation in the bulk is varied from $10^{4}$ - $10^{7}$ and $K^B_N=10^{24}$ (\#/m${^3}$s) is used. An energy density threshold of $E_T=10^7$ J/m$^3$ is adopted. To ensure that most newly nucleated grains that are indeed growing, $r_{nuc}$ = 1.3$r_0$ is taken.
\begin{figure}[H]

\subfloat[]{\hspace{0.5cm} \includegraphics[height=0.21\textheight]{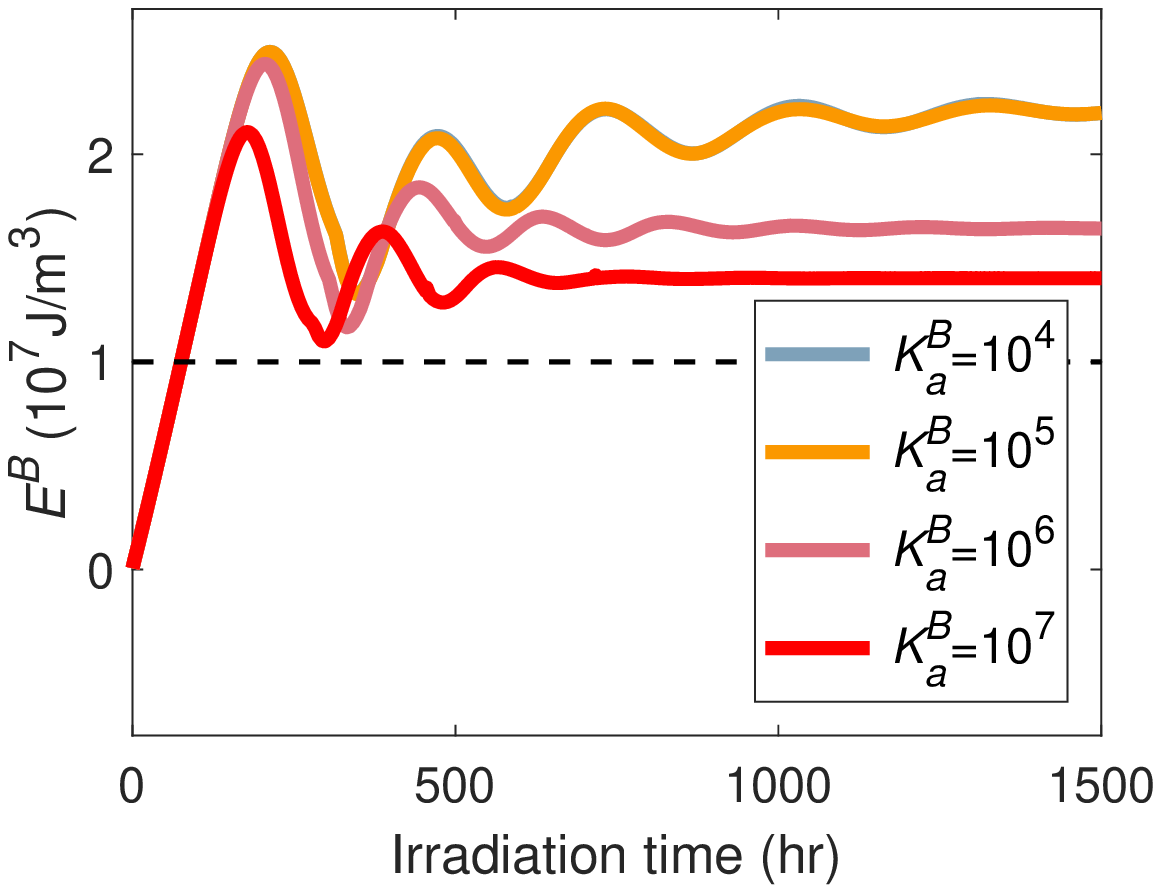}} \hspace{0.01cm}
\subfloat[]{\includegraphics[height=0.21\textheight]{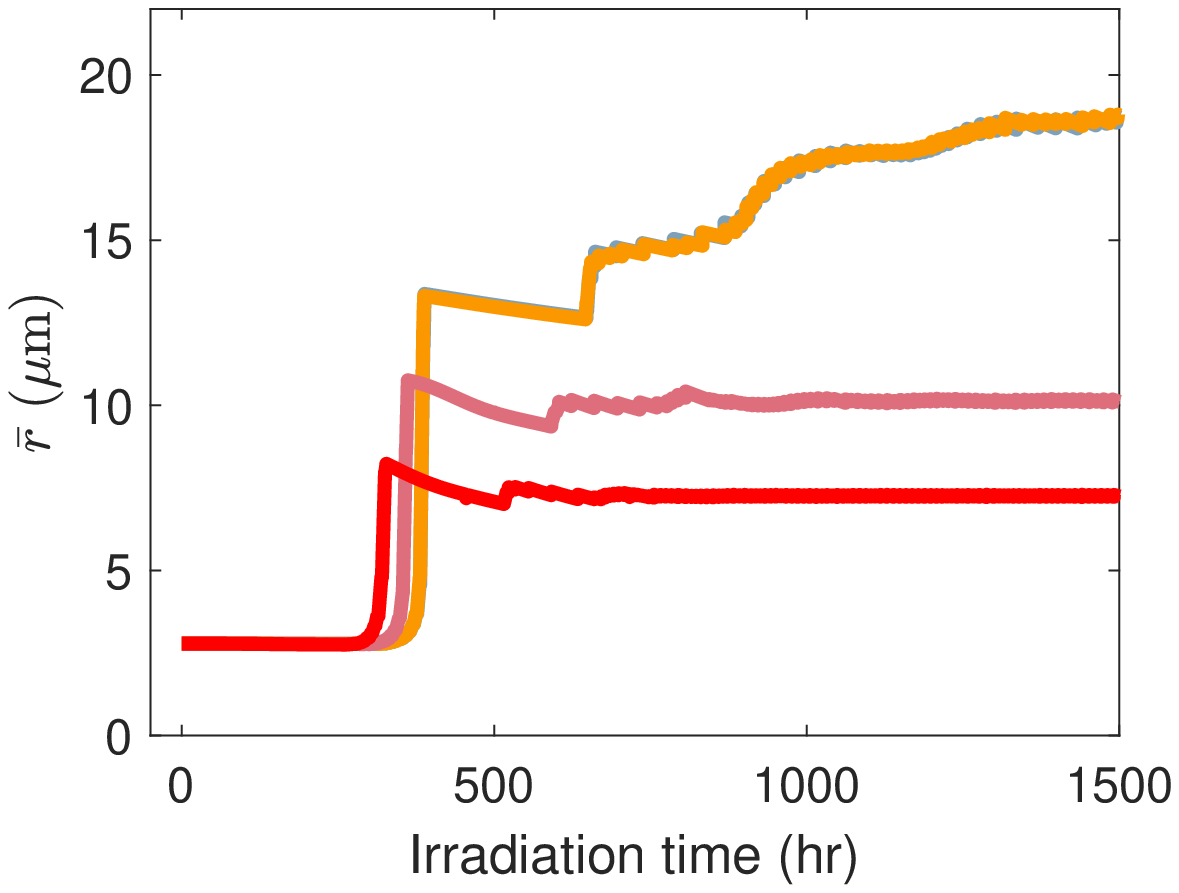}} 
\vspace{0.01cm}
\subfloat[]{\hspace{0.05cm} \includegraphics[height=0.21\textheight]{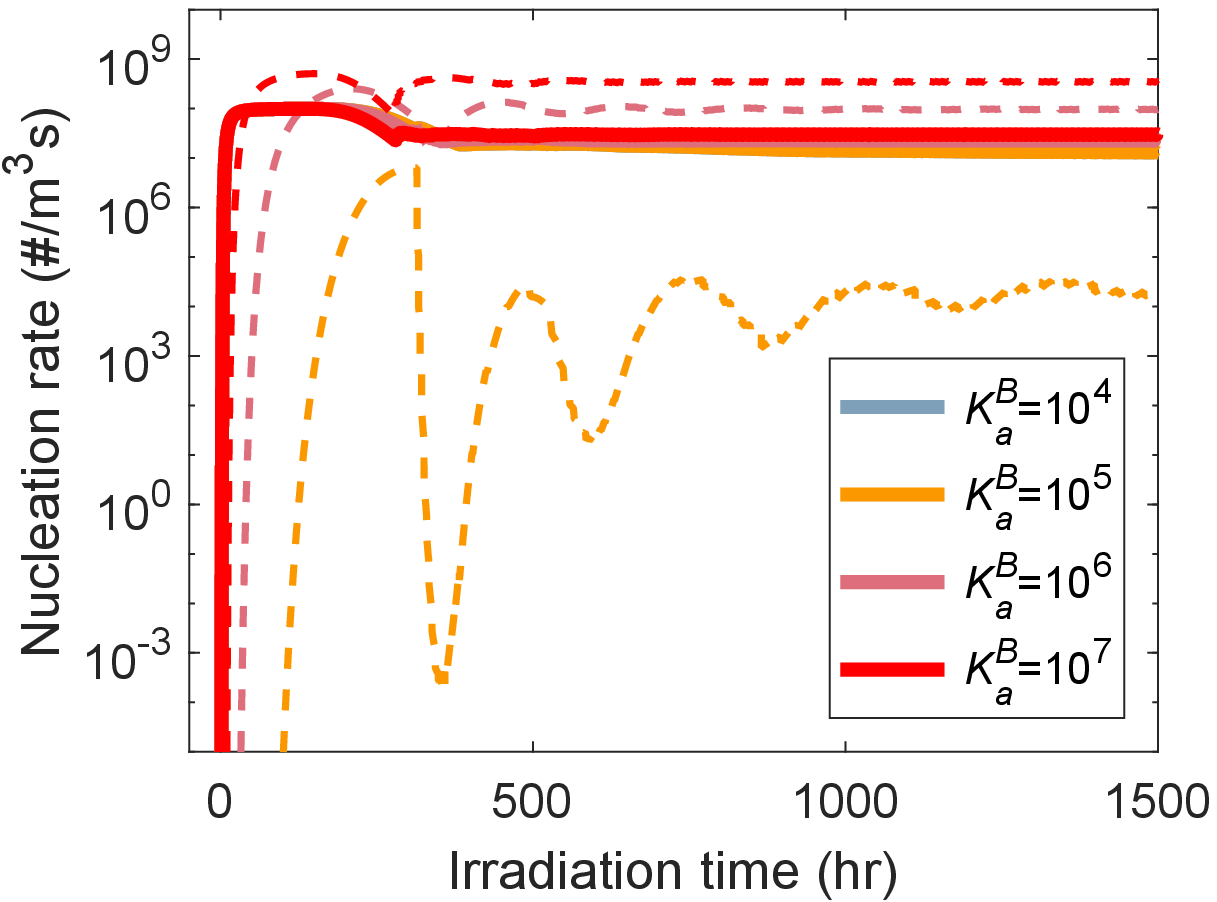}}\hspace{0.2cm}
\subfloat[]{\includegraphics[height=0.21\textheight]{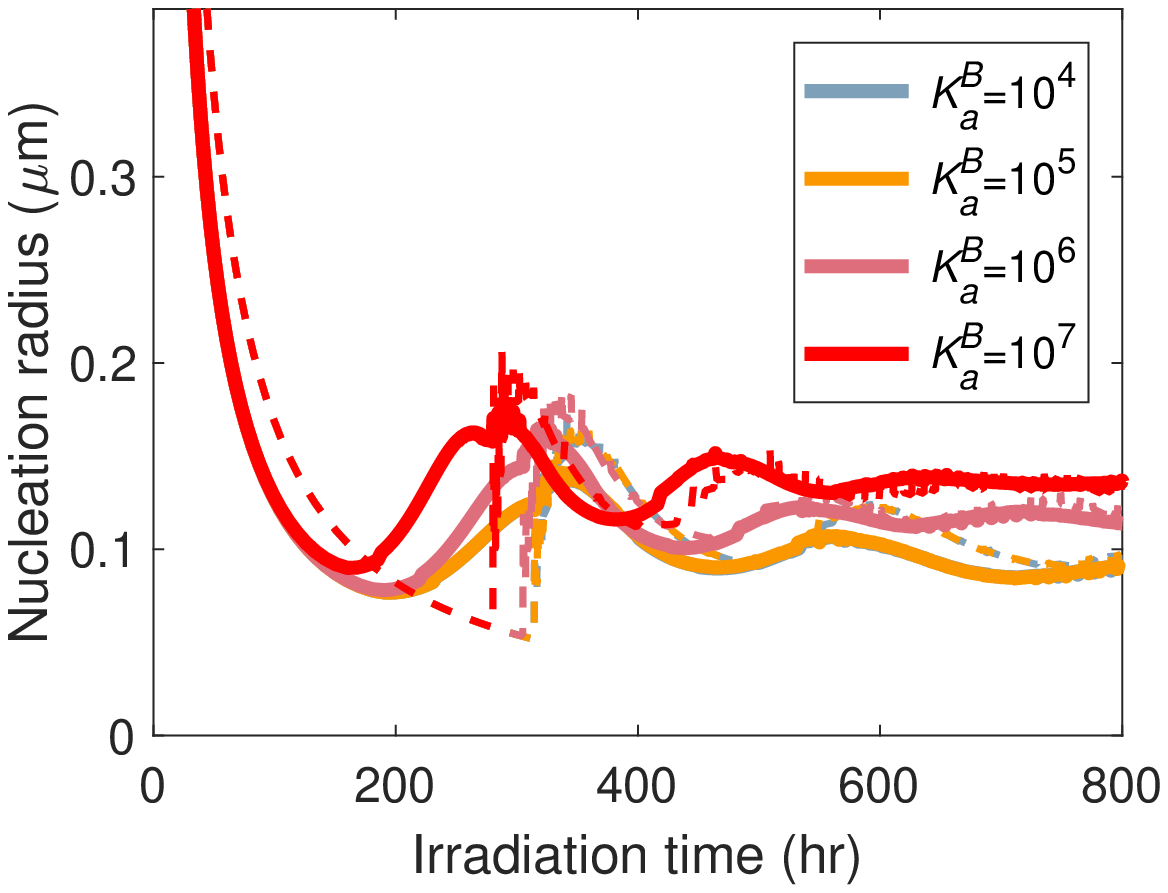}}
\caption[]{Influence of parameter $K_a^B$ on the evolution of (a) the average bulk stored energy density $E^B$, (b) the average grain radius $\bar{r}$, (c) the necklace-type nucleation rate (solid lines) and bulk type nucleation rate (dashed lines) and (d) the nucleation radius for necklace-type nucleation (solid lines) and bulk-type nucleation (dashed lines) for simulations of neutron-induced recrystallization at 1100 \textdegree C.}
\label{fig:loopunf2}
\end{figure}

Figure~\ref{fig:loopunf2} depicts the effects of bulk nucleation on the microstructural evolution. At this temperature, the effects of bulk nucleation on the average grain size and defect energy become visible for values of $K^B_a$ larger than $10^{5}$ ($K^B_a=10^4$ overlaps with $10^5$ in the figure). For these values, the nucleation rate in the bulk is in the same order of, or exceeds, the necklace-type nucleation rate, see Figure~\ref{fig:loopunf2}c. The total nucleation rate increases, which leads to a faster replacement of the original grains (visible from the first dip in $E^B$) and to a shorter cycle for the cyclic recrystallization. The equilibrium value for $E^B$ drops, as well as the average grain size $\bar{r}$. At $K^B_a=10^5$, the bulk nucleation rate strongly depends on $E^B$ (indirectly, through the calculation of $E_{act}^B$), while for  $K^B_a=10^7$, the bulk nucleation barrier becomes so small that the nucleation rate does not depend on the stored energy density anymore.

\paragraph{Evolution of individual nuclei} \mbox{}\\
It is investigated whether the nucleation spot (in the bulk or at the grain boundaries) has a significant effect on the evolution of a nucleus. For this, the simulation at 1100 \textdegree C and $K^B_a=10^7$ is used. During the simulation, a single grain in the bulk and a single grain at the grain boundary are nucleated at specific irradiation times. The grains represent a negligible microstructural volume and therefore do not affect the evolution of the other representative grains. They are not allowed to merge with any other grains, so that the development of their size and stored defect energy can be traced.

\begin{figure}[H]
%\hspace{0.13cm}
%\subfloat[]{\includegraphics[height=0.22\textheight]{Fig11a_inkscape.eps}} 
%\subfloat[]{\includegraphics[height=0.22\textheight]{Fig11b_inkscape.eps}} %\vspace{0.01cm}
%\hfil
%\subfloat[]{\includegraphics[height=0.22\textheight]{NecklaceBulk6_Radius_lin_450.eps}} 
%\subfloat[]{\hspace{0.25cm} \includegraphics[height=0.22\textheight]{NecklaceBulk6_Energy_lin_450.eps}} 
%\hfil
\includegraphics[width=\textwidth]{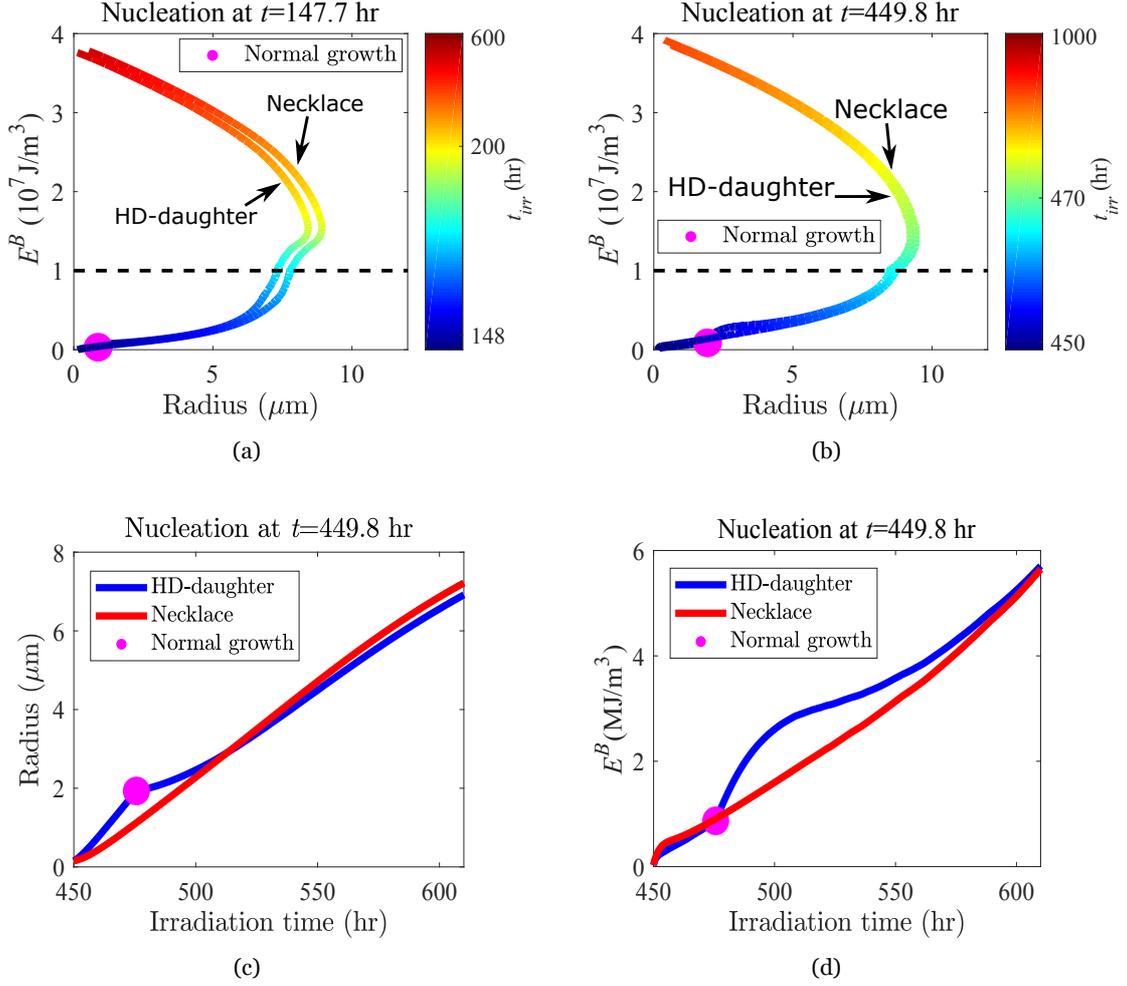}
\caption[]{Simultaneous evolution of the radius and bulk stored energy density of an HD daughter grain and of a necklace-nucleated grain that are formed after (a) 150 hours and after (b) 450 hours of irradiation; evolution of the (c) grain radius and (d) the bulk stored energy density $E^B$ in the grains that were formed after 450 hours of irradiation. The dot indicates the start of normal growth. All results are taken at an irradiation temperature of 1100 \textdegree C with $K^B_a=10^7$.}
\label{fig:ind}
\end{figure}Figure~\ref{fig:ind} shows the results of the evolution of the individual HD daughter grain and necklace nucleated grain for initiation after 147.7 hours and 449.8 hours of irradiation. The evolution paths for both grains are quite similar, but not the same. The necklace nucleated grain reaches a slightly higher maximum radius than the HD daughter grain; this was the case for each of the initiation times that were simulated approximately every 50 hours, ranging from 50 to 500 hours). The point at which the HD daughter grain resumes normal growth is marked. Figure ~\ref{fig:ind}c and ~\ref{fig:ind}d explain why the $(r,E^B)$-paths in Figure~\ref{fig:ind}b do not overlap: the HD daughter starts with a much faster growth than the necklace nucleated grain. After resuming normal growth, the defect energy accumulates rapidly. The paths in the $(r,E^B)$ graphs are quite similar but not fully synchronized in time, as is clearified by Figure~\ref{fig:ind}c and d.

\section{Conclusions and discussion}
A model was developed for neutron-induced recrystallization using cluster dynamics and a mean-field recrystallization model. The model includes the incorporation of interstitial loops to become part of the dislocation network and nucleation in both the bulk and at the grain boundaries. In the recrystallization model, distinction was made between the growth mechanism of a bulk nucleus and of a grain boundary nucleus. The effects of nucleation in the bulk on the evolution of the overall microstructural behavior and the evolution of individual grains were investigated. The predicted microstructural properties were used to make qualitative predictions for the evolution of the hardness during the neutron-induced recrystallization.\\
\\
With the loop incorporation included, a cyclic neutron-induced recrystallization process is systematically predicted. This cyclic behaviour is apparent from the evolution of the bulk stored energy density, as well as from the evolution of the grain size and grain energy distributions. The period for complete renewal of the grains is clearly temperature-dependent. Over time the microstructure becomes more heterogenous and the periodicity fades out. \\
\\
Based on the concentrations of the vacancy clusters, interstitial clusters and dislocation densities, the increasing hardness was assessed, using a hardness indicator. The type of defects that dominate the hardening in these simulations were identified as dislocations. The Hall-Petch effect was shown to be negligible in comparison to the hardening due to obstacles at the grain interior; the evolution of the integrity of the grain boundaries was not taken into account here. The obtained simulation results on the evolution of the hardness call for more experimental results on the evolution of this quantity under irradiation, also for low doses. The inclusion of loop incorporation as a mechanism into the model leads to high values for the dislocation density and to a (faster) saturation of the defect concentrations in a single grain, when recrystallization is not included (as was shown in cluster dynamics simulations on a single grain).\\
\\
The effect of bulk nucleation on the evolution of the microstructural properties has been studied. Depending on the nucleation barrier, the bulk nucleation rate may heavily depend on $E^B$ for large barriers, or only very little if the nucleation barrier is very low. A higher total nucleation rate will lead to a lower equilibrium value for $E^B$, a smaller average grain size and a shorter period for cyclic recrystallization. Nuclei that form and grow in the bulk initially grow faster and accumulate less damage; afterwards they evolve quite similarly as necklace nucleated grains. Because of their faster initial growth, they accumulate a somewhat higher bulk stored defect energy density, which leads to a lower maximum grain size.\\
\\
One way to increase the lifetime of the divertor monoblocks could be by means of a heat treatment. With the developed model, it can be explored what the effect of possible heat treatments on the microstructural evolution could be. By linking the mechanical properties to the microstructural properties, an optimal microstructure may be determined. This model can then be applied to identify the conditions under which such a microstructure can be formed.

\section{Data availability}
The data generated or analysed in this study are available upon simple request.

\appendix 
\section{Solution procedure}\label{sec:appimp}
The following steps are taken to solve the grain growth and nucleation model. 

\paragraph{Nucleation}\mbox{}\\
The procedure is explained for necklace nucleation. For bulk nucleation, a similar procedure is used. The main differences are pointed out. 
\begin{enumerate}
\setlength \itemsep{0.01cm}
\item The volume averaged total HEM-energy densities $E^{HD}$ and $E^{LD}$ and the volume averaged bulk stored energy densities of the HEMs, $E^{B,HD}$ and $E^{B,LD}$ are calculated, as well as the total energy density $E_k$ and bulk energy density $E^B_k$ for each grain $k$.
	\begin{itemize} 
	\item Here, $E^B_k=\sum_{n=1}^{N_{max}}\bigg[ C_{I_n,k}E^f_{I_n} + C_{V_n,k}E^f_{V_n}\bigg]+ \mu b^2 \rho_k/2  - 		TS_k$, where $C_{I_n,k}$ and $C_{V_n,k}$ are the concentrations of clusters $I_n$ and $V_n$ in grain $k$.
	\end{itemize}
\item The mobile surface fractions $\phi^{LD}$ and $\phi^{HD}$ are calculated: 
	\begin{itemize} \setlength \itemsep{0.01cm}
	\item The volume fraction of LD-grains: $f^{LD} = V^{LD} / (V^{LD} + V^{HD})$. 
	\item The mobile surface fraction of LD grains is given by $\phi^{LD} = 1-\bigg( f^{LD} \bigg)^{2/3}$. The condition of volume conservation (the total volume transferring from LD to HD should equal the total volume that transfers from HD to LD) is met by using:\\
$\phi^{HD}=-\phi^{LD}\big[\sum_{j\in LD} r_j^2 N_j \Delta E^{HD}_j \big]/\big[\sum_{i\in HD} r_i^2 N_i \Delta E^{LD}_i\big]$. 
	\item The maximum value of $\phi^{HD}$ is limited to 1. In that case $\phi^{LD}$ is calculated based on $\phi^{HD}$, such that volume conservation is met.
	\end{itemize}
\item $E^B = \phi^{LD} E^{B,HD} + (1-\phi^{LD}) E^{B,LD}$ and $E = \phi^{LD} E^{HD} + (1-\phi^{LD}) E^{LD}$.
\item The numerical derivatives $\frac{dE^B}{dt}$ and $\frac{dE^{B,HD}}{dt}$ are calculated using a linear least-squares fit to approximate the derivative: \\
$dE^B/dt \approx Cov(t,E^B) / Var(t)=\big[\sum_{i=1}^n (t_i - \bar{t})(E^B_i - \bar{E^B})\big] / \big[\sum_{i=1}^n (t_i - \bar{t})^2 \big]$, where the results of the last hour of irradiation are used.
\item The radius and activation energy for necklace nucleation are determined: 
	\begin{itemize} \setlength \itemsep{0.01cm}
	\item All roots of to the polynomial $d \Delta E^A/dt$=0 are determined. The largest real solution is $r_L$.
	\item The static solution $r^*=3\gamma_b/(2(E^B-E^B_0))$ is calculated. 
	\item The candidate solution for the nucleus radius is $r^A_{nuc}=1.01\mathrm{x }max(r_L,r^*)$. 
	\item If $d \Delta E^A(r^A_{nuc})/dt<0$, then the candidate solution is stable. If this is the case, then $E^{A}_{act}=\Delta E (r^A_{nuc})$. If not, then $\dot{N}^A=0$.
	\end{itemize}
\item The nucleation surface area, which consists of all HD/HD and HD/LD-grain boundaries, is $A_{nuc} =2 \pi \big(\sum_{i\in HD} r_i^2 N_i + \phi^{LD} \sum_{j\in LD} r_j^2 N_j \big)$.
\item The amount of new nuclei in this time increment is: \\ $N = \dot{N}^A \Delta t= K_N^A A_{nuc}\mathrm{exp}\big( -1/k_BT [E^A_{act} + Q_{GB}/N_A] \big) \Delta t$.
\item The total nucleation volume $V^A_{nuc}=4 \pi N \big( r^A_{nuc}\big)^3$. This nucleation volume is delivered by the other grains. The new radii for the HD- and LD-grains are respectively:
	\begin{itemize} \setlength \itemsep{0.01cm}
	\item $r^{HD} := r^{HD} \bigg( 1 - [\phi^{LD} V^A_{nuc}]/[\sum_{i\in HD} \frac{4 \pi}{3} N_i r_i^3 ]\bigg)^{1/3}$;
	\item $r^{LD} := r^{LD} \bigg( 1 - [(1-\phi^{LD}) V^A_{nuc}]/[\sum_{j\in LD} \frac{4 \pi}{3} N_j r_j^3 ]\bigg)^{1/3}$.
	\end{itemize}
\item For bulk nucleation, the procedure is similar, with main differences:
	\begin{itemize}
	\item $V_{nuc} = \sum_{i\in HD} \frac{4 \pi}{3} r_i^3 N_i$; 
	\item The new radii for the HD-grains change a second time during the nucleation procedure, to become: 
	\\ $r^{HD} := r^{HD} \bigg( 1 - [ V^V_{nuc}]/[\sum_{i\in HD} \frac{4 \pi}{3}       N_i r_i^3 ]\bigg)^{1/3}$.
	\end{itemize}
\item The nucleated grains are added to the LD-set. If necessary, the merging procedure is used. In this procedure, $E^S$ and $E^B$ are calculated for all $q$ grains that exist within the HEM (that are of the same type) and for the grain that will be added to the set during the procedure. For each of the $q+1$ candidates for merging, the square difference with all the other grains, $(E^S-E^{S'})^2 + (E^B-E^{B'})^2$, is calculated. The grains with the smallest square difference are merged. The grain properties are volume averaged. 
\item If any HD daughter grain satisfies $r>\bar{r}_{HD}/4$, it is subsequently treated as a normal grain.
\end{enumerate}

\begin{figure}[ht!]
\includegraphics[width=0.5\textwidth]{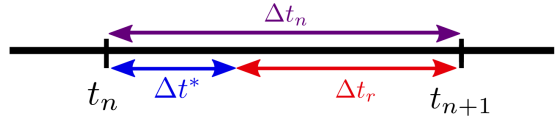}
\caption[]{Schematic illustration of the time step subincrementation, in case of vanishing grains.}
\label{fig:vangr}
\end{figure}

\paragraph{Grain Growth}\mbox{}
\begin{enumerate}
\setlength \itemsep{0.01cm}
\item The HD-daughter grains that grow only at the expense of the HD-interior, are kept separate. For all the other grains, steps 2-6 are performed.
\item The energy density difference $\Delta E^{HEM}_k = E^{HEM} - E_k$ for each grain is computed.
\item The mobile surface fractions $\phi^{LD}$ and $\phi^{HD}$ are computed. 
\item The volume changes per grain $\Delta V^{HEM}_k$ are computed. For LD-grains that shrink with respect to their own HEM, $\Delta V_j^{LD} = \frac{\sum_{\Delta E^{LD}_j>0}r_j^2 \Delta E_j^{LD} N_j}{\sum_{\Delta E^{LD}_j<0}r_j^2 \Delta E_j^{LD} N_j} (1-\gamma^{LD}) 4 \pi m \Delta t r_j^2 \Delta E_j^{LD}$ is used, and likewise for the HD-grains that shrink with respect to their own HEM.
\item Based on the volume changes, the concentrations and the new grain radii are updated. The new concentrations are $C(t_{i+1}) = C(t_i) \bigg( 1 - \frac{\Delta V_k^{growth}}{V_k(t_{i+1})}\bigg) + C^{eq}\frac{\Delta V_k^{growth}}{V_k(t_{i+1})}$, where $C(t)$ is the concentration of a certain defect of a certain grain $k$, $\Delta V_k^{growth}$ is the defect-free volume that is added to the grain and $V_k(t_i)$ is the grain volume at time $t_i$. The thermal equilibrium concentrations for vacancy and interstitial defects are determined from their formation energies: $C^{eq} = \mathrm{exp}(-E^f/k_BT)$.
\item In case a grain volume becomes negative during the increment, the following procedure is used: 
\begin{itemize}
\item A subincrement for the time step is used, for which the grain exactly vanishes: $\Delta t^* = \Delta t_i \frac{V(t_{n})}{V(t_{n}) - V(t_{n+1})}$, see Figure~\ref{fig:vangr}. 
\item Steps 2-6 of the grain growth routine are repeated, starting from $t_n$, using the subincrement $\Delta t^{*}$. 
\item To complete the increment, steps 2-5 of the grain growth routine are repeated, starting from $t=t_i+\Delta t^{*}$, using a time step size $\Delta t_r$  (see Figure~\ref{fig:vangr}). In case another grain volume becomes negative during the remainder of the increment, the time interval $\Delta t_r$  is further subdivided using the described procedure. 
\end{itemize}
\item Finally, each HD-daughter grain $k$ interacts with the HD-HEM, leading to a volume change of: $\Delta V_k = 4 \pi r_k^2 m \Delta E^{HD}_k \Delta t$. Using this, the new grain radius and new defect concentrations are calculated.
\end{enumerate}

\section{Modelling details} 
\subsection{Rate coefficient details} \label{app:rc}
The formulas for computation of each rate coefficient can be found in Table \ref{tab:ratecoef}. The parameter values that have been used can be found in Table \ref{tab:parval}.

\begin{table*}[ht!]
		\caption[]{Rate coefficients. The superscript `+ ` denotes absorption of a point defect and the subscript `-' denotes emission.} 
	\label{tab:ratecoef}
\centering
	\begin{tabular}[ht!]{l| >{$}r<{$} |  >{$}r<{$} }

	 \multicolumn{2}{c}{\bf{Dislocation loop \(\mathbf{ I_n} \) }} \\ 
	%\text{Dislocation loop } I_n 				&																	&\\
&	\textit{Absorption rate}					&	\textit{Emission rate} 													\\ \hline
Interstitial & 	\alpha_n^+ = 2 \pi r_{I_n}Z_{I_n}^I D_I 		&  \alpha_n^- =  2 \pi r_{I_{n-1}}Z_{I_{n-1}}^I D_I \text{ exp } ( - E_{I_n-I}^b / k_B T ) / V_{at}	 		\\ Vacancy    & k_{I_n+V}^+ = 2 \pi r_{I_n}Z_{I_n}^V D_V		& k_{I_{n-1}-V}^- =  2 \pi r_{I_{n-1}}Z_{I_{n-1}}^V D_V \text{ exp } ( - E_{I_n-V}^b / k_B T ) / V_{at}		 \\
 VI-recombination&	k^+_{I+V} = 4 \pi r_{IV} (D_I + D_V)   		& 																 \\
	\multicolumn{2}{c}{\bf{Vacancy cluster  \(\mathbf{V_n} \)}} \\
&	\textit{Absorption rate}					&	\textit{Emission rate} 													\\ \hline
Interstitial & 	k_{V_n+I}^+ = 4 \pi r_{V_n} D_I 			& 	-															\\
Vacancy &	\gamma_n^+ = 4 \pi r_{V_n} D_V			& \gamma_n^- = 4 \pi r_{V_{n-1}} D_V \text{ exp } \big( -E^b_{V_n-V} / k_B T\big)	/ V_{at}			 \\
	\end{tabular}
\end{table*}

\begin{table*}[ht!]	
	\caption[]{Rate coefficients related to the strengths of the sinks (grain boundaries and dislocations).} 
	\begin{tabular}[t]{ >{$}r<{$}|  l}
\multicolumn{2}{c}{\textbf{Dislocation sink \( \rho_D \)}} \\ \hline
	k_{D+I}^+ = \rho_D Z_D^I D_I 				& Interstitial \\									
	k_{D+V}^+ = \rho_D Z_D^{V} D_{V} 			& Vacancy \\
	& \\
	\multicolumn{2}{c}{\textbf{Grain boundary sink}}  \\  \hline
 	k_{S+I}^+ = 3 S_I^{sk} D_I / r_{grain} & Interstitial\\
 	k_{S+V}^+ = 3 S_V^{sk} D_V / r_{grain} & Vacancy \\
\end{tabular}
\end{table*}
\begin{table*}
\caption[]{Expressions for the grain boundary sink strength, the formation energies and binding energies of the defect clusters, the diffusion coefficients of the mobile defects and the bias factors for the interaction between mobile defects and self-interstitial clusters. For the simulations, $r_p=2b$ is assumed \cite{Li2012}.} 
	\begin{tabular}[t]{ >{$}l<{$}     >{$}l<{$}l}
	\multicolumn{2}{c}{\textbf{Sink strength coefficient for GB-sink strength calculation (based on \cite{Bullough1980})}}\\
	\bigg( S_{I}^{sk} \bigg)^2	&	=  \frac{1}{D_I} \bigg[2 \alpha^+_1 C_{I_1} + \sum_{n=2}^{N_I-1} \alpha^+_n C_{I_n} + \sum_{n=1}^{N_V}  k^+_{V_n+I} C_{V_n} \bigg] + \rho Z_D^I  \\
	\bigg(S_{V}^{sk} \bigg)^2 &		=    \frac{1}{D_V} \bigg[ 2 \gamma^+_1 C_{V_1}  + \sum_{n=2}^{N_V-1}  \gamma^+_n C_{V_n} + \sum_{n=1}^{N_I} k^+_{I_n+V} C_{I_n} \bigg] + \rho Z_D^V\\
 	&	 \\
	\multicolumn{2}{c}{\textbf{Binding energy (using the capillarity approximation)}} \\
	E^f_{I_n} & = E^f_{I_{n-1}}+E^f_{I} -E^b_{I_{n}-I} \text{ (by definition)} \\
	E_{I_n-I}^b & =E_I^f + \frac{E_{I_2}^b-E_I^f}{2^{2/3}-1} \big[ n^{2/3} - (n-1)^{2/3} \big]  \\
	E_{I_n-V}^b & =E_V^f + \frac{E_{I}^f-E_{I_2}^b}{2^{2/3}-1} \big[ n^{2/3} - (n-1)^{2/3} \big] \\
	E_{V_n-V}^b &=E_V^f + \frac{E_{V_2}^b-E_V^f}{2^{2/3}-1} \big[ n^{2/3} - (n-1)^{2/3} \big]  \\
	
	& \\
 	\multicolumn{2}{c}{\textbf{Diffusion coefficients}} \\
	\multicolumn{2}{c}{$D_I=D_{I_0} \text{ exp }(-E_I^m/k_B T)$} \\
	\multicolumn{2}{c}{$D_V=D_{V_0} \text{ exp }(-E_V^m/k_B T)$} \\
	& \\
	\multicolumn{2}{c}{\textbf{Dislocation bias factor}}\\
	\multicolumn{2}{c}{$Z_{I_n}^{I}=Z_D^{I} \text{max} \big[ \frac{2 \pi}{\ln(8 r_{I_n}/r_p)},1 \big] $} \\	
	\multicolumn{2}{c}{$Z_{I_n}^{V}=Z_D^{V} \text{max} \big[ \frac{2 \pi}{\ln(8 r_{I_n}/r_p) },1\big] $} \\		
\end{tabular}
\end{table*}

\subsection{Entropy of point defect clusters} \label{app:ent}
The point defect clusters contribute to the entropy of mixing of the system. As long as the defect density can be called dilute, the sum of all the individual contributions of each defect type to the mixing entropy can be taken as \cite{Olander1972}.
\begin{equation}
S=k_B \ln \Omega
\end{equation} $S$ is the entropy density (J/(m$^3$K)) and $\Omega$ is the number of possible configurations of the defects in the material, with: 

\begin{align}
\ln \Omega = \sum_{m} \Omega_m = & \sum_m \bigg[ \ln \frac{m^{N(m)}(\frac{N_S}{m})!}{(\frac{N_S}{m}-N(m))! (N(m))!} \bigg] \\ \nonumber
= &\sum_{m} \bigg[ N(m) \ln m + \frac{N_S}{m} \ln (\frac{N_S}{m}) \\ \nonumber
& - [\frac{N_S}{m} - N(m)] \ln [ \frac{N_S}{m} - N(m)]  \\ \nonumber
& - N(m) \ln [N(m)] \bigg]
\end{align}
Here  $m$ denotes the number of point defects in the cluster, $N_S$ is the number of lattice sites that is availabe in a unit volume. If the unit volume is 1 m$^3$, then $N_S=1/V_{at}$ and if the concentrations are in atomic units, then $N_S=1$. Lastly, $N (m)$ is the defect distribution. 

\subsection{Model parameters} \label{app:par}
\begin{table}[H]
\centering
\caption[]{Parameter values}
\label{tab:parval}
	\begin{tabular}{>{$}l<{$} | l      |       l                |            l              |  r}
	\textbf{Parameter}		&	\textbf{Unit}	& \textbf{Value} & \textbf{Description}		&	\textbf{Source}						\\ \hline
	a_0			&	nm	&	0.31652	& Lattice parameter 			& \cite{Lassner1999} 				\\
	\gamma_b 		&	J/m$^2$	&	0.869		& GB-surface energy		& \cite{Lopez2015} 	\\
	D^{GB}_0		&	m$^2$/s	&	0.27\e{-4}	& Self-diffusivity & Estimated, \\
	& & & along grain boundaries&  using \cite{Lassner1999}		\\	
	\delta			& 	nm 	&	1		& GB thickness			& \cite{Favre2013}   	\\	
	D_{I_0}		&	m$^2$/s	&	8.77\e{-8}	& SIA-diffusivity & \cite{Faney2013}	          \\
	D_{V_0}		&	m$^2$/s	&	177\e{-8}	& Vacancy diffusivity& \cite{Faney2013}		\\	
	E^f_I 			&	eV	&	9.466		& Formation energy SIA		& \cite{Li2012}		\\
	E^f_V			&	eV	&	3.80		& Formation energy vacancy	& \cite{Li2012}		\\
	E^b_{I_2}		&	eV	&	2.12		& Binding energy SIA-SIA		& \cite{Li2012}		\\
	E^f_{V_2}		&	eV	&	0.6559	& Binding energy V-V		& \cite{Li2012}		\\
	E^m_I 		&	eV	&	0.013		& SIA migration energy 		& \cite{Faney2013}		\\
	E^m_V 		&	eV	&	1.66		& Vacancy migration energy 	& \cite{Faney2013}		\\
	K_a^A   		& 	- 	&	1\e{8}		& Nucleation activation energy reduction & -			\\	
	K_m 			&	- 	& 	1.8\e{4}	& GB mobility parameter 		& -				\\
	K_{m_0}		& 	- 	& 	25		& GB mobility parameter with pinning &			-	\\
	K_N^A         		&	\#/m$^2$s &	3.16\e{17}	& Necklace nucleation rate constant	& -				\\
	K_N^V         		&	\#/m$^3$s &	1\e{24}	& Bulk nucleation rate constant	& -				\\	
	M 			& - 		&  	3.06   		& Taylor factor 			& \cite{Stoller2000} 	\\
	\mu			&	Pa 	&	161\e{9} 	& Shear modulus ($T=0 \degree C$)			&  \cite{Lassner1999}	\\	
	N_{max}	          &	- 	& 	100		& Maximum cluster size		& -				\\	
	Q^{GB}		&	J/mol	&	4\e{5}	& Activation energy for GB mobility	& \cite{Lassner1999}				\\
	r_{IV}			& nm   	&       0.465               & Recombination radius		& \cite{Li2012}                 \\       
	\rho^{eq} 		&  m$^{-2}$ & 	10$^9$	&Equilibrium dislocation density & - \\
	Z_D^I 		&	- 	& 	1.2		& SIA-dislocation bias		&  \cite{Li2012}		\\
	Z_D^V		& 	- 	&	1		& V-dislocation bias			&	\cite{Li2012}		\\	
	\end{tabular}
	%}
\end{table}

\begin{table}[ht!]
\centering
    \begin{tabular}{| l | l | l | l | }
    \hline
	$T$ (K)				&	$G_0$ ($\#$ defects/atom s)	&	$S_I$ 		&	$S_V$	 	\\ \hline
	300					&	4.3\e{-8}				&	2.20		&	1.63 		\\ \hline
	1025					&	3.3\e{-8} 				&	2.50		&	1.86 		\\ \hline
	2050					&	3.1\e{-8} 				&	2.17		&	2.42 		\\ 
        \hline
    \end{tabular}
    \caption[]{Parameter values for the displacement damage production rate $G_0$ and for the power law exponents $S_I$ and $S_V$ at different temperatures, from \cite{Mannheim2018}, based on MD-results \cite{Setyawan2015}. Linear interpolation is used to obtain the parameter values at other temperatures.}
    \label{tab:ownpowerlaw}
\end{table}

\subsection{Dislocation evolution details} \label{app:dis}
Here, the expressions governing the evolution of the dislocation density (Equation~\ref{eq:gen3}), based on \cite{Stoller1990}, are given. The density of the Bardeen-Herring sources $S_{BH}$ depends on the pinned dislocation density $\rho_p$ as $S_{BH}=\bigg( \rho_p/3 \bigg)^{1.5}$. The pinned dislocation density is assumed to be given by $\rho_p=0.1 \rho$, following \cite{Stoller1990}. The dislocations climb with a velocity $v_{cl}$, which is given by $v_{cl}=\frac{2 \pi}{b \ln{R/r_c}} \bigg( Z^I_D D_I C_I - Z^V_D D_V (C_V - C_V^D) \bigg)$. Here $Z^V_D$ and $Z^I_D$ are the same dislocation biases as in the equations for the rate coefficients, see Appendices~\ref{app:par} and \ref{app:rc}, $R$ and $r_c$ are the dislocations outer and core radii ($R/r_c=2 \pi$ is used here) and $C_V^D$ is the equilibrium concentrations of vacancies nearby a dislocation. $C_V^D$ depends on the equilibrium vacancy concentration in the bulk $C_V^{eq}$ and on the internal stress $\sigma$ due to immobilized dislocations, by: $C_V^D = C_V^0 \exp{\bigg( \frac{\sigma V_{at}}{k_B T}\bigg)}$ \cite{Was2007}, with $\sigma = A \mu b \sqrt{\rho_p}$, where $A=0.4$ is a geometric parameter and $\mu$ is the shear modulus. Finally, the dislocations can travel a distance $d_{cl} = \big( \pi \rho \big)^{-1/2}$ before they annihilate.

\newpage
\subsection{Cluster evolution details} \label{app:cd}
The full set of cluster dynamics equations is given below, adapted from \cite{Li2012}. At every cluster growth term, a factor $1-f_n(\rho_t)$ is added to account for the incorporation of prismatic loops, after \cite{Jourdan2015}.
\subparagraph{The full Cluster Dynamics equations}
\begin{dgroup*}
\begin{dmath*}
\frac{dC_I}{dt} = 	G_I  + k_{I_2+V}^+C_{I_2} C_V + 2 \alpha_2^- C_{I_2}   + \sum_{n=3}^{N_I} \alpha_{n}^- C_{I_{n}} - k_{I+V}^+ (C_I C_V - C_I^{eq} C_V^{eq})  - 2 \alpha_1^+C_I^2   - \sum_{n=2}^{N_I} \alpha_{n}^+ C_I C_{I_{n}}  - \sum_{n=2}^{N_V} k_{V_n+I}^+ C_I C_{V_n} - (k_{D+I}^++k_{S+I}^+) C_I  
%%%%%%%%%%%
\end{dmath*}
\begin{dmath*}
 \frac{dC_{I_2}}{dt} =   	 G_{I_2} + \alpha_1^+ C_I^2 + k_{{I_3}+ V}^+ C_{I_3} C_V + \alpha_3^- C_{I_3}  + k^-_{I-V} C_I   - \alpha_2^- C_{I_2}  - \alpha_2^+ C_{I_2} C_I - k^-_{I_2-V} C_{I_2} - k^+_{I_2+V}C_{I_2}C_V
%%%%%%
\end{dmath*}
\begin{dmath*}
\frac{dC_{I_{n}}}{dt}_{3\le n\le N_I-2}=		  G_{I_n}+  \alpha_{n+1}^- C_{I_{n+1}} + (1-f_n(\rho_t)) \big[ \alpha_{n-1}^+ C_I C_{I_{n-1}}
+ k_{I_{n-1}-V}^- C_{I_{n-1}} \big] - \alpha_n^{+} C_I C_{I_n} + k_{I_{n+1}+V}^+ C_V C_{I_{n+1}}  - \alpha_{n}^- C_{I_{n}}      -k_{I_{n}+V}^{+} C_V C_{I_n}  - k_{I_n-V}^- C_{I_n}
\end{dmath*}
\begin{dmath*}
\frac{dC_{I_{N_{I}-1}}}{dt} =	G_{I_{N_I-1}} +\alpha_{N_I}^- C_{I_{N_I}} + (1-f_n(\rho_t)) \big[ \alpha_{N_I-2}^+ C_I C_{I_{N_I-2}} + k_{I_{N_I-2}-V}^- C_{I_{N_I-2}} \big] - \alpha_{N_I-1}^+ C_I C_{I_{N_I-1}}  - \alpha_{N_I-1}^- C_{I_{N_I-1}} -k_{I_{N_I-1}+V}^{+} C_V C_{I_{N_I-1}}   + k_{I_{N_I}+V}^+ C_V C_{I_{N_I}} 	- k_{I_{N_I-1}-V}^- C_{I_{N_I-1}} 
\end{dmath*}
\begin{dmath*}
\frac{dC_{I_{N_{I}}}}{dt}= 		 G_{I_{N_I}}+ (1-f_n(\rho_t)) \big[ \alpha_{N_I-1}^+ C_I C_{I_{N_I-1}} 	+ k_{I_{N_I-1}-V}^- C_{I_{N_I-1}} - \alpha_{N_I}^- C_{I_{N_I}} -k_{I_{N_I}+V}^{+} C_V C_{I_{N_I}} 
\end{dmath*}
\begin{dmath*}
 \frac{dC_V}{dt} =   G_V + 2 \gamma_2^- C_{V_2} +k^+_{V_2+I}C_{V_2}C_I  + \sum_{n=3}^{N_V} \gamma_{n}^- C_{V_{n}} 
 + \sum_{n=2}^{N_{I+1}} k_{I_{n-1}-V}^- C_{I_{n-1}} - k_{I+V}^+ (C_I C_V - C_I^{eq} C_V^{eq})  - 2 \gamma_1^+ C_V^2 - \sum_{n=2}^{N_V} \gamma_n^+ C_V C_{V_n} 
 - \sum_{n=2}^{N_I}  k_{I_n+V}^+ C_V C_{I_n}  -    (k_{D+V}^++k_{S+V}^+)C_V 
\end{dmath*}
\begin{dmath*}
 \frac{dC_{V_{n}}}{dt}_{2 \le n \le N_V-2} = 	G_{V_n} +   k_{V_{n+1}+I}^+ C_I C_{V_{n+1}} + \gamma_{n+1}^-  C_{V_{n+1}}   - k_{V_n+I}^+ C_I C_{V_n} 
								+ \gamma_{n-1}^+ C_V  C_{V_{n-1}}
 							   - \gamma_n^- C_{V_n} - \gamma_n^+ C_V C_{V_{n}} 
\end{dmath*}
\begin{dmath*}
 \frac{dC_{V_{N_V-1}}}{dt} =  		  G_{N_{N_V-1}} + k_{V_{N_V}+I}^+ C_I C_{V_{N_V}} + \gamma_{N_V}^-  C_{V_{N_V}} 
								+ \gamma_{N_V-2}^+ C_V  C_{V_{N_V-2}}
 							  - k_{V_{N_V-1}+1}^+ C_I C_{V_{N_V-1}}  - \gamma_{N_V-1}^- C_{V_{N_V-1}}
  							- \gamma_{N_V-1}^+ C_V C_{V_{N_V-1}} 
\end{dmath*}
\begin{dmath*}
 \frac{dC_{V_{N_V}}}{dt} = 			  G_{N_{N_V}} + \gamma_{N_V-1}^+ C_V  C_{V_{N_V-1}} 
 							  - k_{V_{N_V}+I}^+ C_I C_{V_{N_V}}  - \gamma_{N_V}^- C_{V_{N_V}} 
\end{dmath*}
\end{dgroup*}

\section{Additional results}
\subsection{Influence of the damage rate on the cluster dynamics results}
\label{app:damrate}

\begin{figure}[H]
%\hspace{0.5 cm}
\subfloat[]{\includegraphics[height=0.22\textheight]{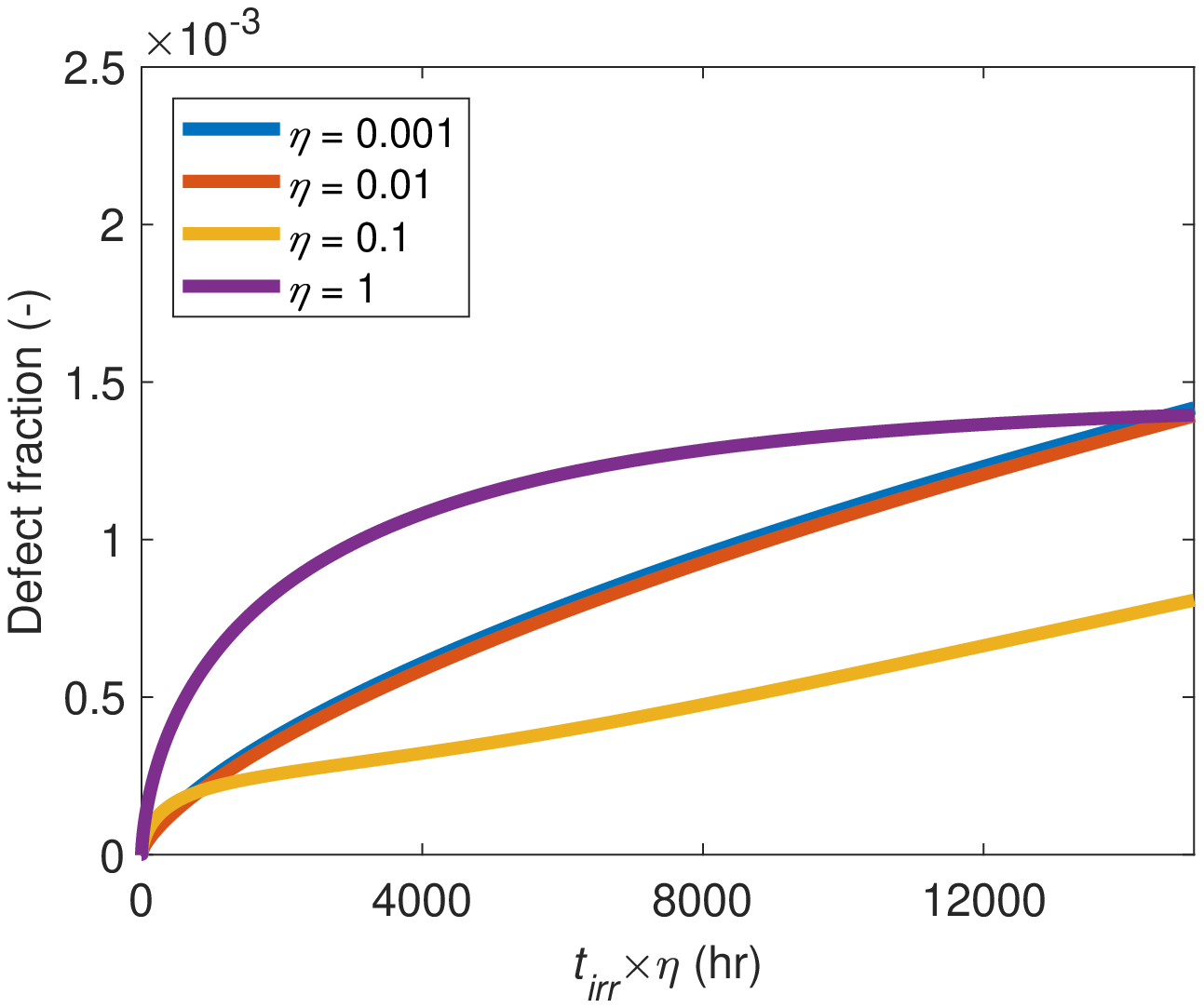}} \hspace{0.1cm}
\subfloat[]{\includegraphics[height=0.22\textheight]{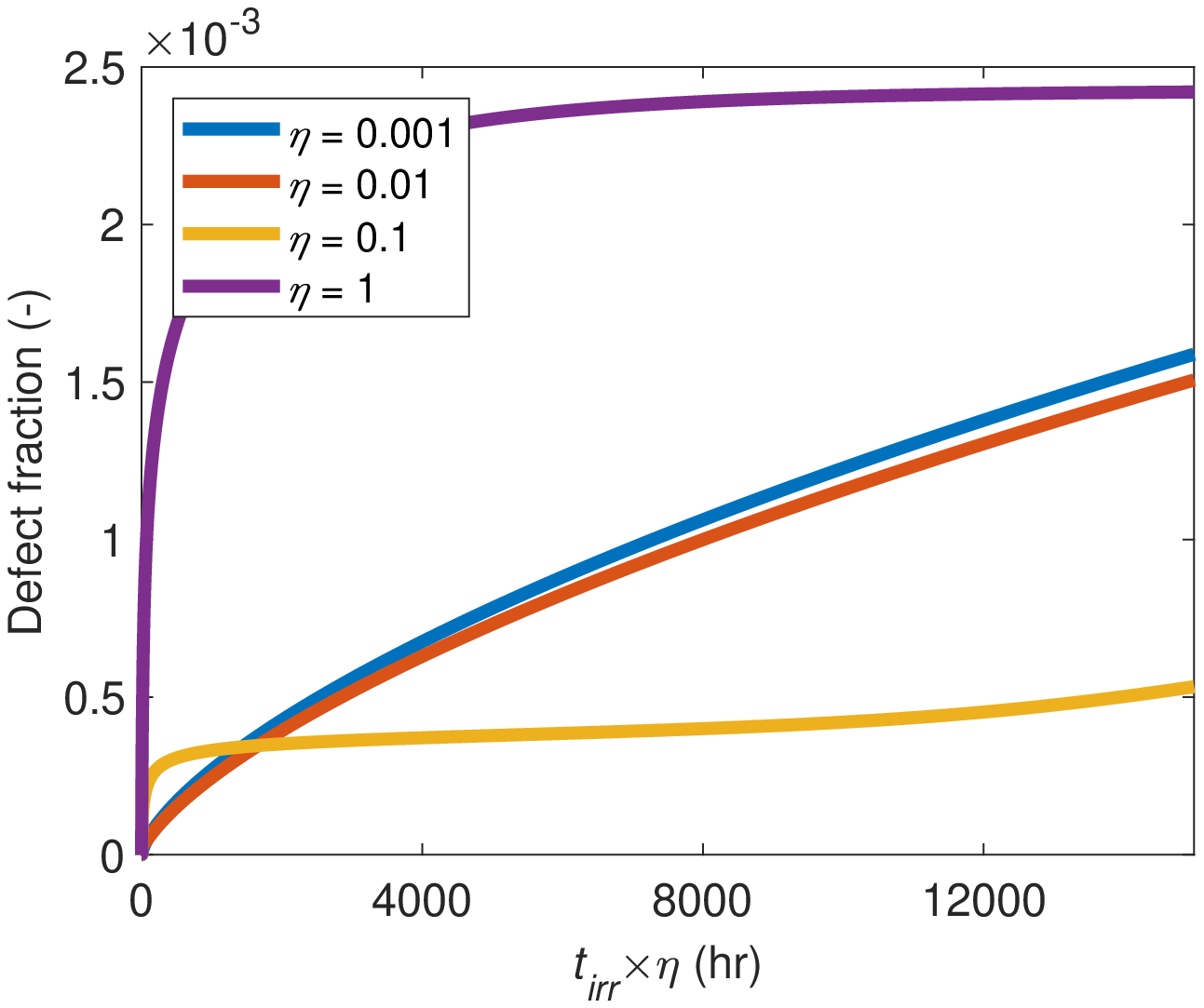}} \vspace{0.1cm}
\subfloat[]{\includegraphics[height=0.22\textheight]{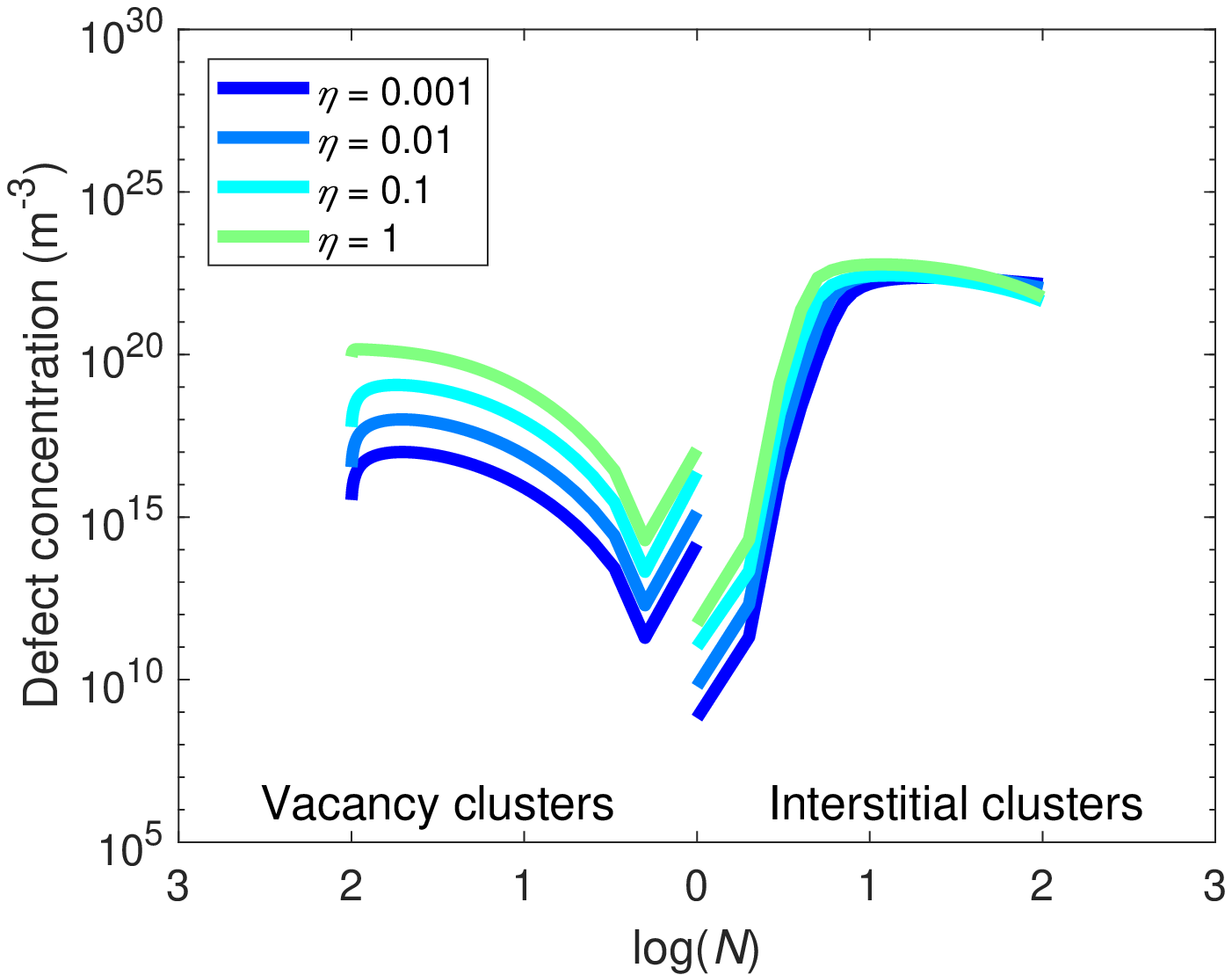}}\hspace{0.1cm}
\subfloat[]{\includegraphics[height=0.22\textheight]{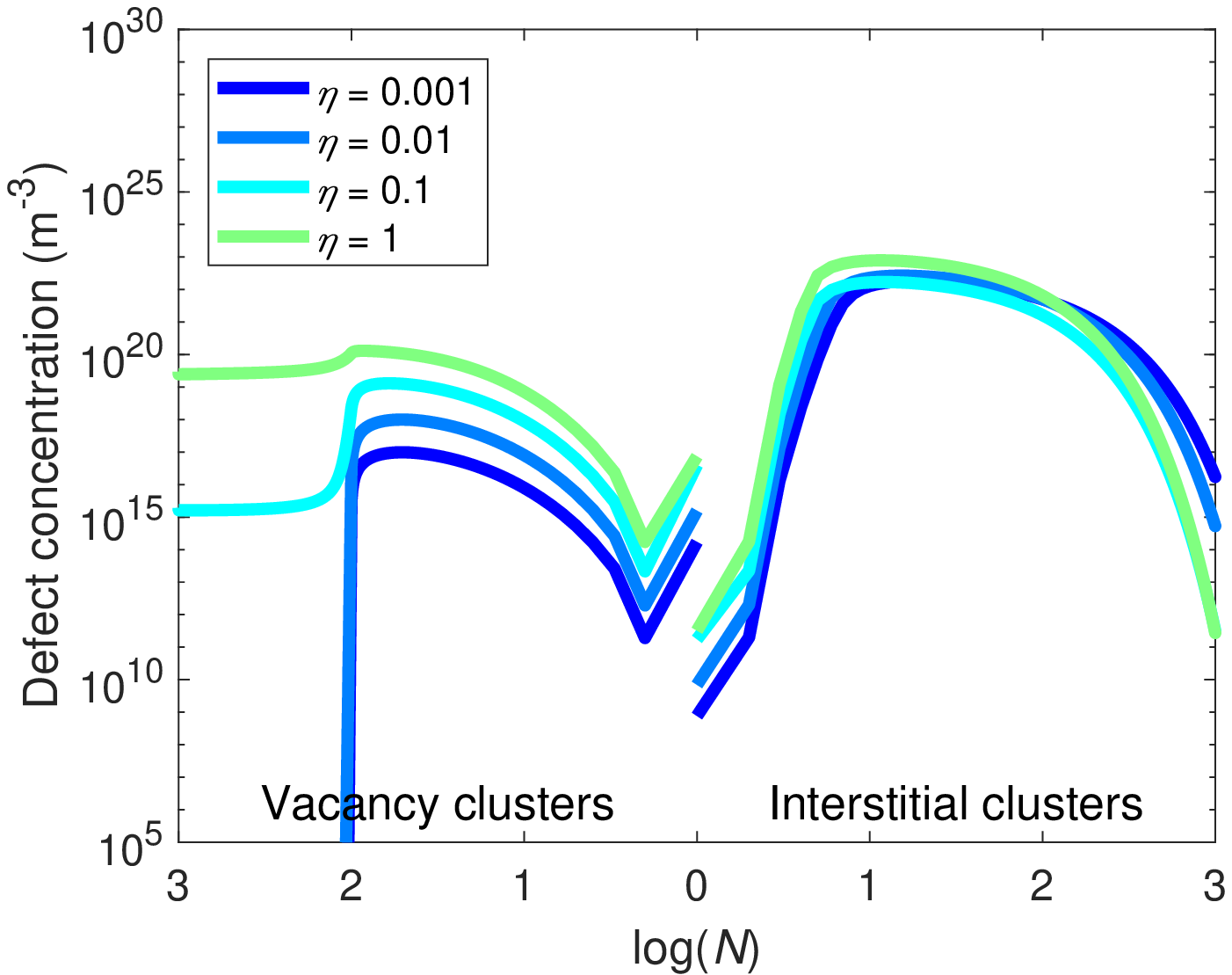}}
\caption[]{Evolution of the atomic defect fraction at an irradiation temperature of $1000$ \textdegree C for (a) $N_{max}$=100 and (b) $N_{max}$=1000. (Note that for the purpose of comparison, the irradiation time is multiplied by $\eta$.) Concentrations of the vacancy clusters and interstitial clusters at $t_{irr} \times \eta = 15000$ hr, for (c) $N_{max}=100$ and (d) $N_{max}=1000$. }
\label{fig:damrate}
\end{figure}

To illustrate the influence of the damage rate on the results produced with the cluster dynamics model, the evolution of the defect fractions and the final concentration distributions are shown for $\eta$ = 0.001, 0.01, 0.1 and 1 and for $N_{max}$ = 100 and 1000, at an irradiation temperature of 1000 \textdegree C, in Figure~\ref{fig:damrate}. For simplicity, loop incorporation is not considered in these simulations. The trends are the same for both values of $N_{max}$. The origin of the differences in the results for the different damage rates is simple, \textit{i.e.}: at low damage rates, it takes longer for the material to receive the same amount of damage. Therefore, the mobile defects have more time to diffuse to sinks (dislocations, grain boundaries). Moreover, vacancy clusters have more time to emit vacancies and SIA-clusters absorb more vacancies. This leads to lower concentrations for the mobile defects, vacancy clusters and the SIA-clusters of medium size. The concentrations of the large SIA-clusters are not significantly affected at this stage. Later on, the lower concentration of medium-sized SIA-clusters leads to considerable absorption of mobile SIAs at large SIA-clusters, for the lowest damage rate ($\eta$=0.001). The differences in defect concentrations that thus form, explain the differences in the evolution of the defect fraction. The defect fraction evolves linearly with the amount of applied damage, diminished by the  annihilation rate at defect sinks. From the results, it was found that for each damage rate, initially, the main defect sinks are vacancies and dislocations. From roughly $\eta \times t_{irr}$=0.1 hr, the main manner in which mobile defects are removed, is by the absorption of vacancies at SIA-clusters.

\subsection{Evolution of the grain size distribution} \label{app:cumdist}

 \begin{figure}[H]
\subfloat[]{\includegraphics[height=0.22\textheight]{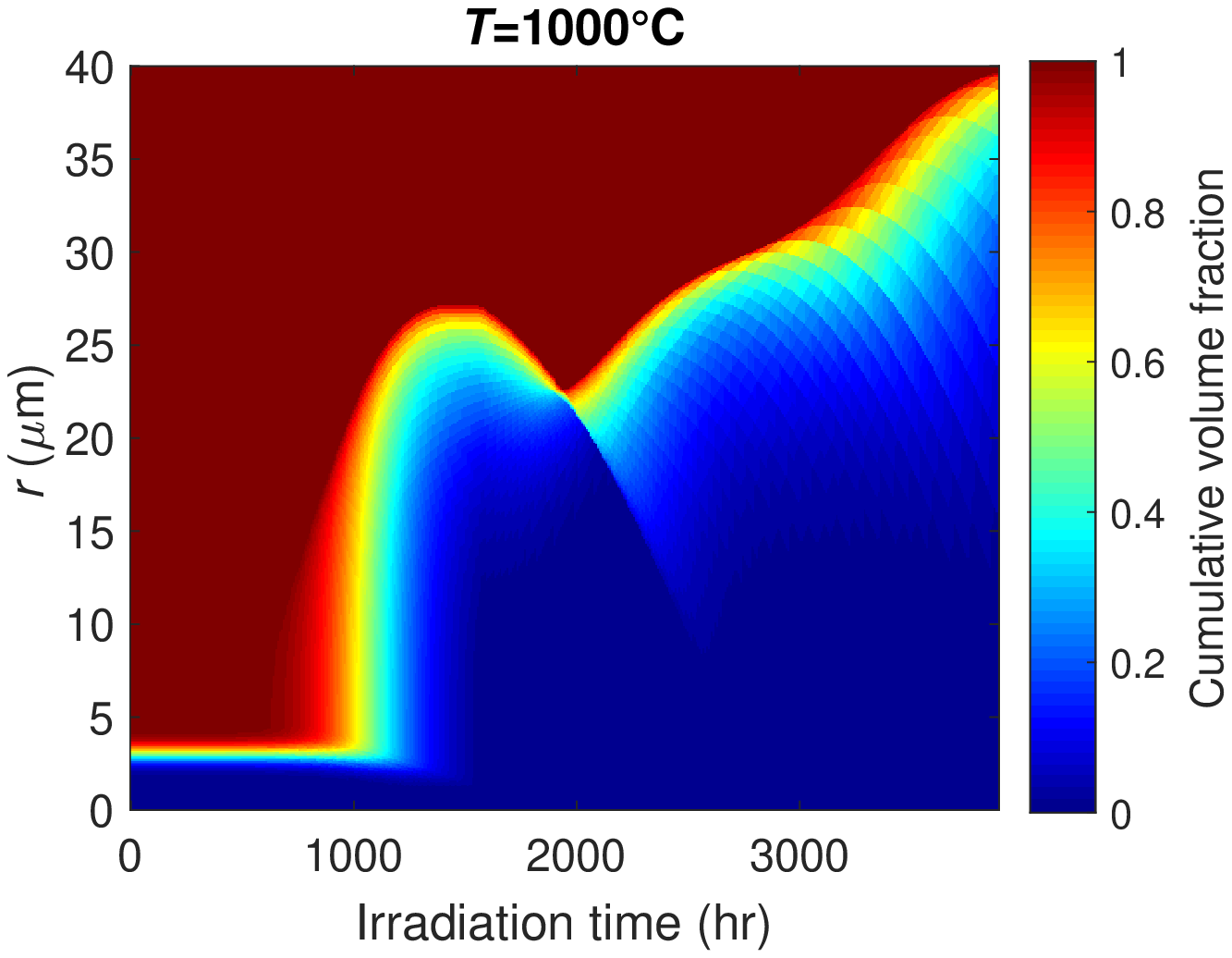}} \hspace{0.1cm}
\subfloat[]{\includegraphics[height=0.22\textheight]{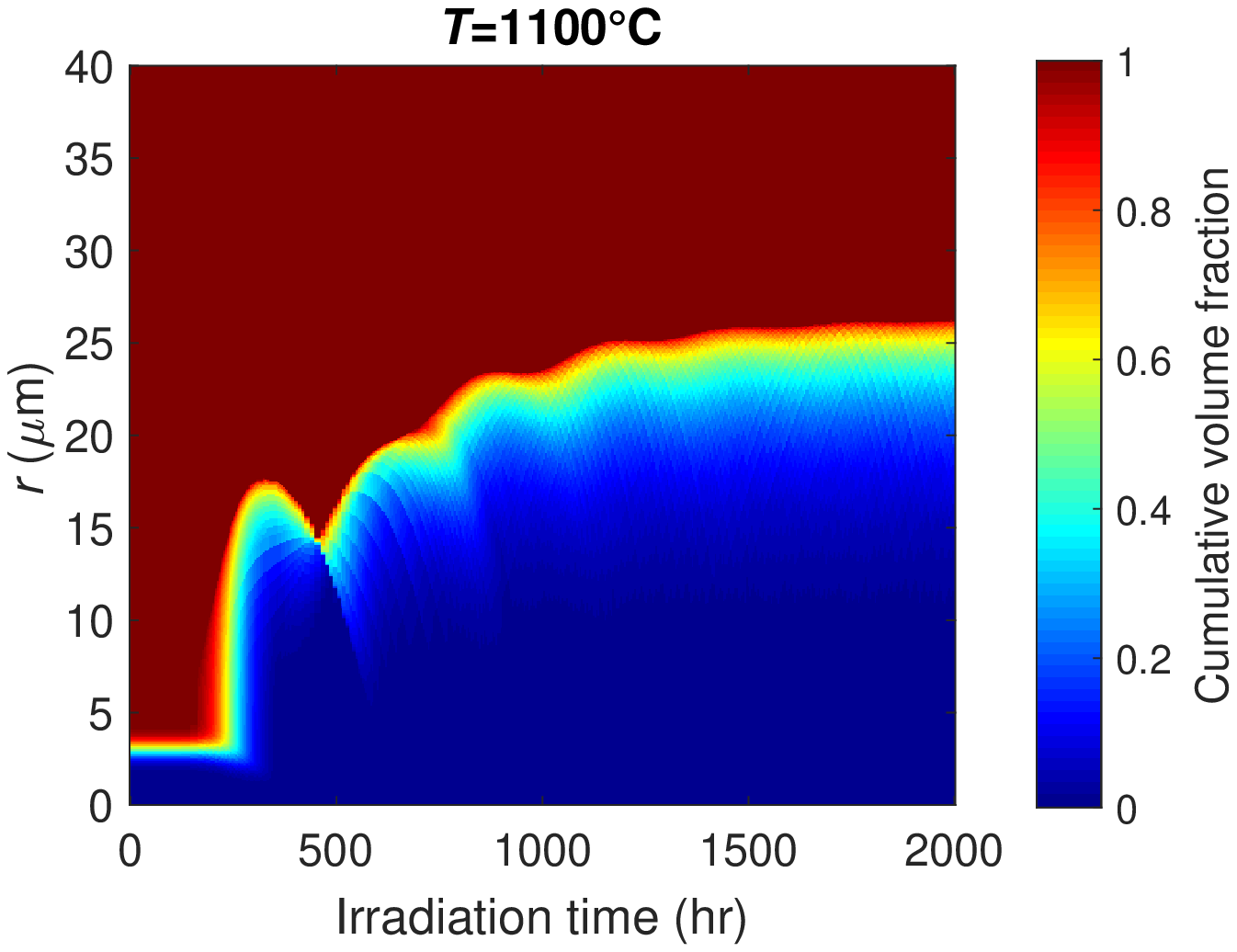}} 
\vspace{0.01cm}
\subfloat[]{\includegraphics[height=0.22\textheight]{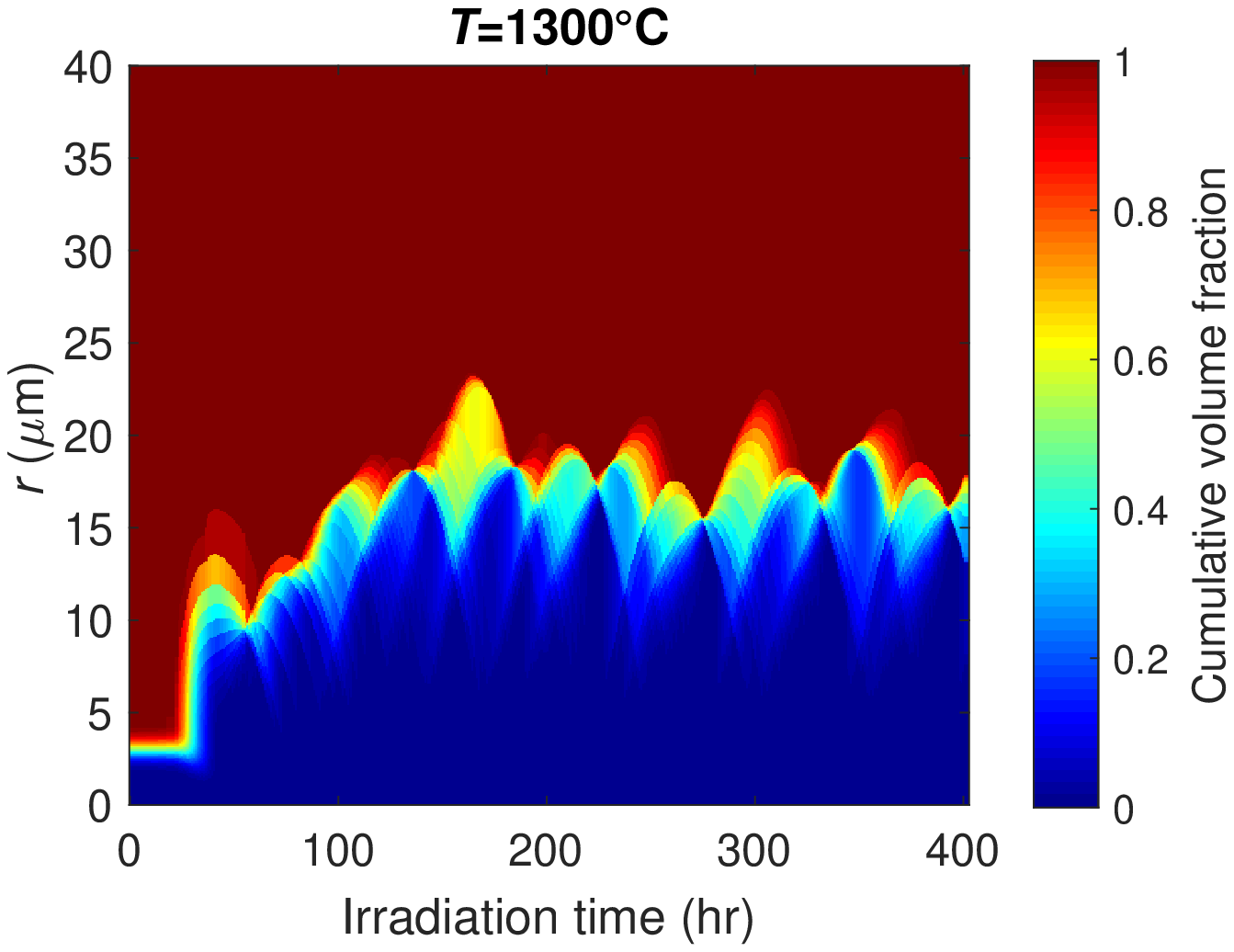}} \hspace{0.1cm}
%\subfloat[]{\includegraphics[height=0.22\textheight]{ImprUnf4_Radmap_cumsum1400.eps}} 
\caption[]{Evolution of the grain size distributions at (a) 1000 \textdegree C, (b) 1100 \textdegree C and (c) 1300 \textdegree C, in terms of cumulative volume fractions.}
\label{fig:gdist}
\end{figure}

 \begin{figure}[H]
\subfloat[]{\includegraphics[height=0.22\textheight]{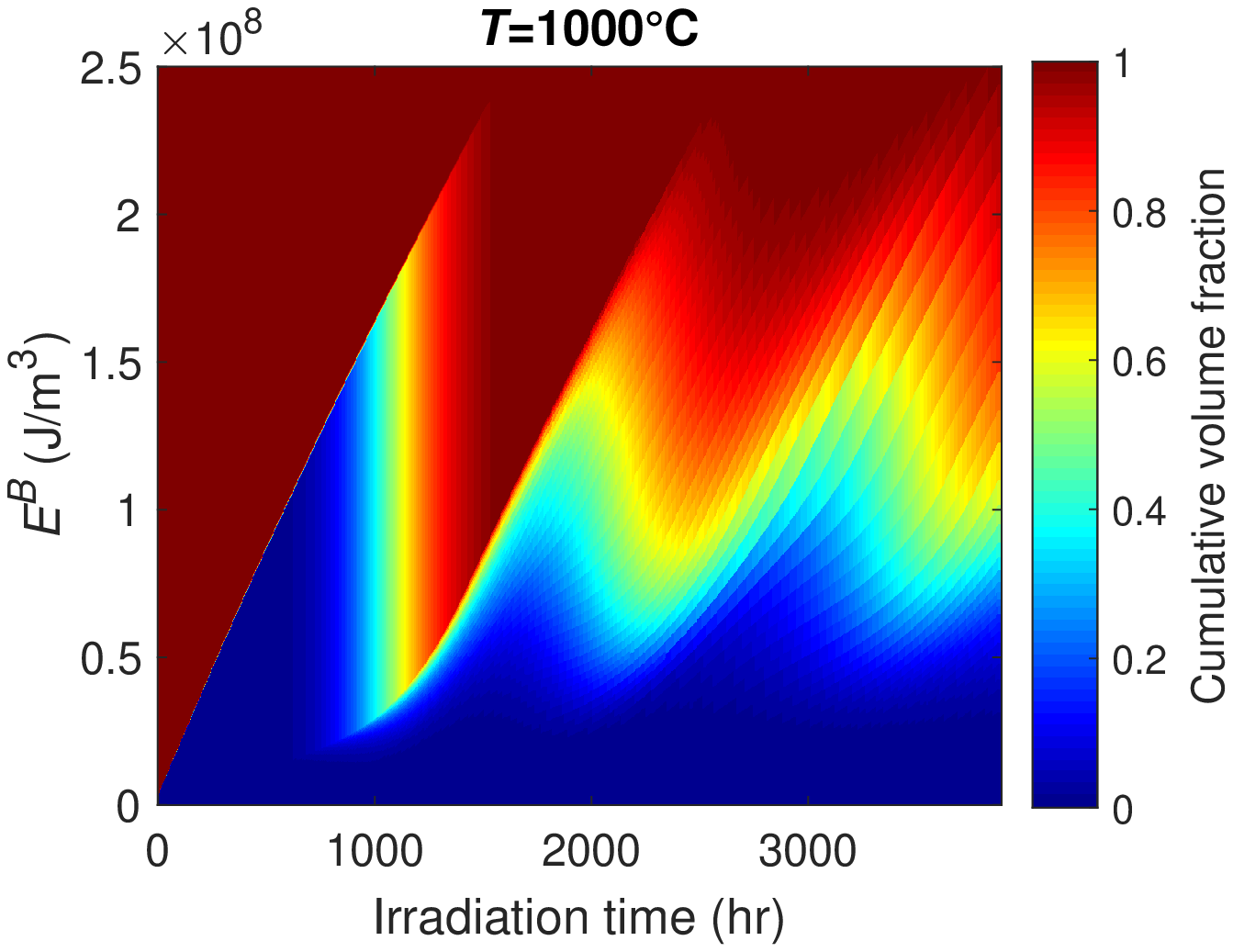}} \hspace{0.1cm}
\subfloat[]{\includegraphics[height=0.22\textheight]{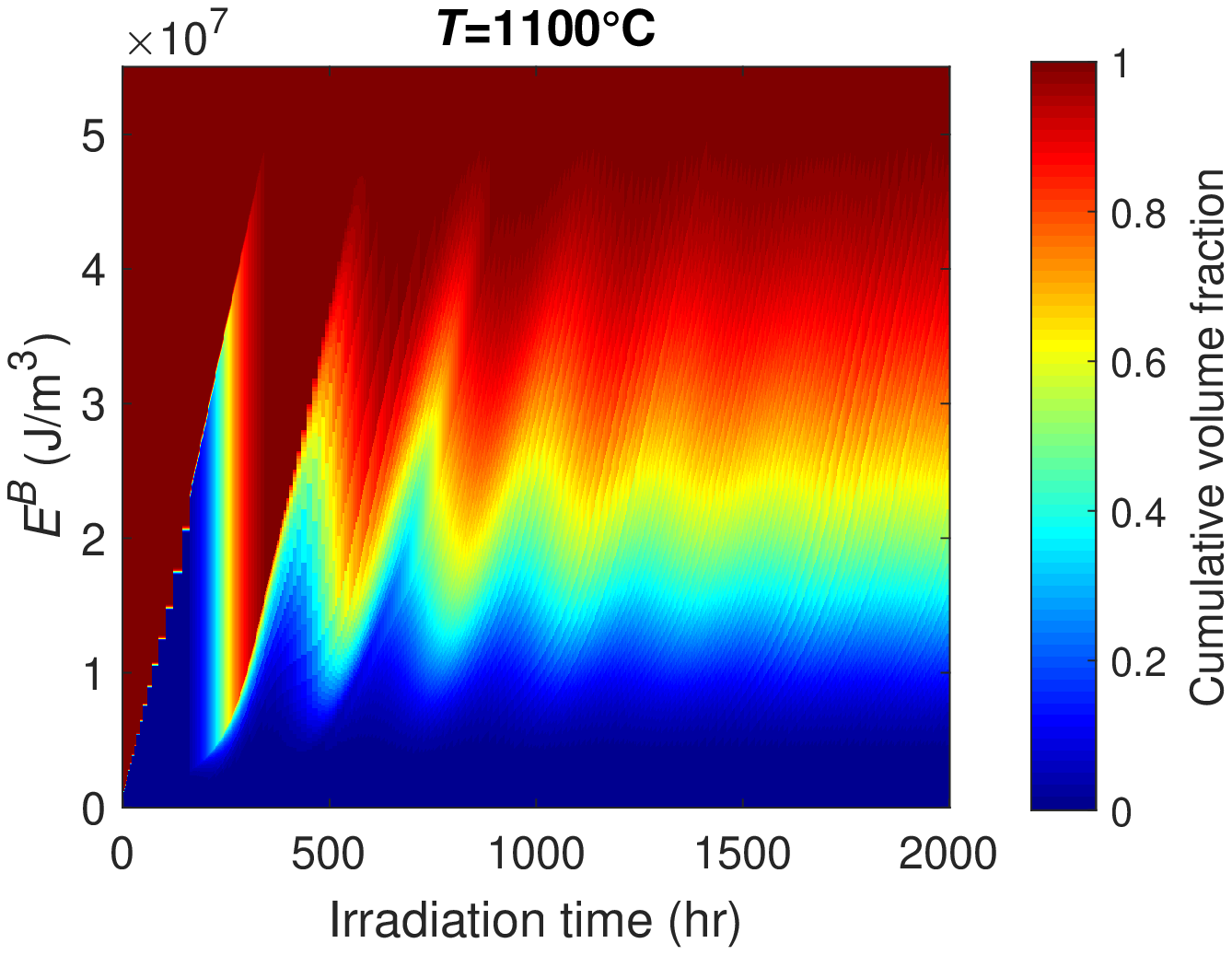}} 
\vspace{0.01cm}
\subfloat[]{\includegraphics[height=0.22\textheight]{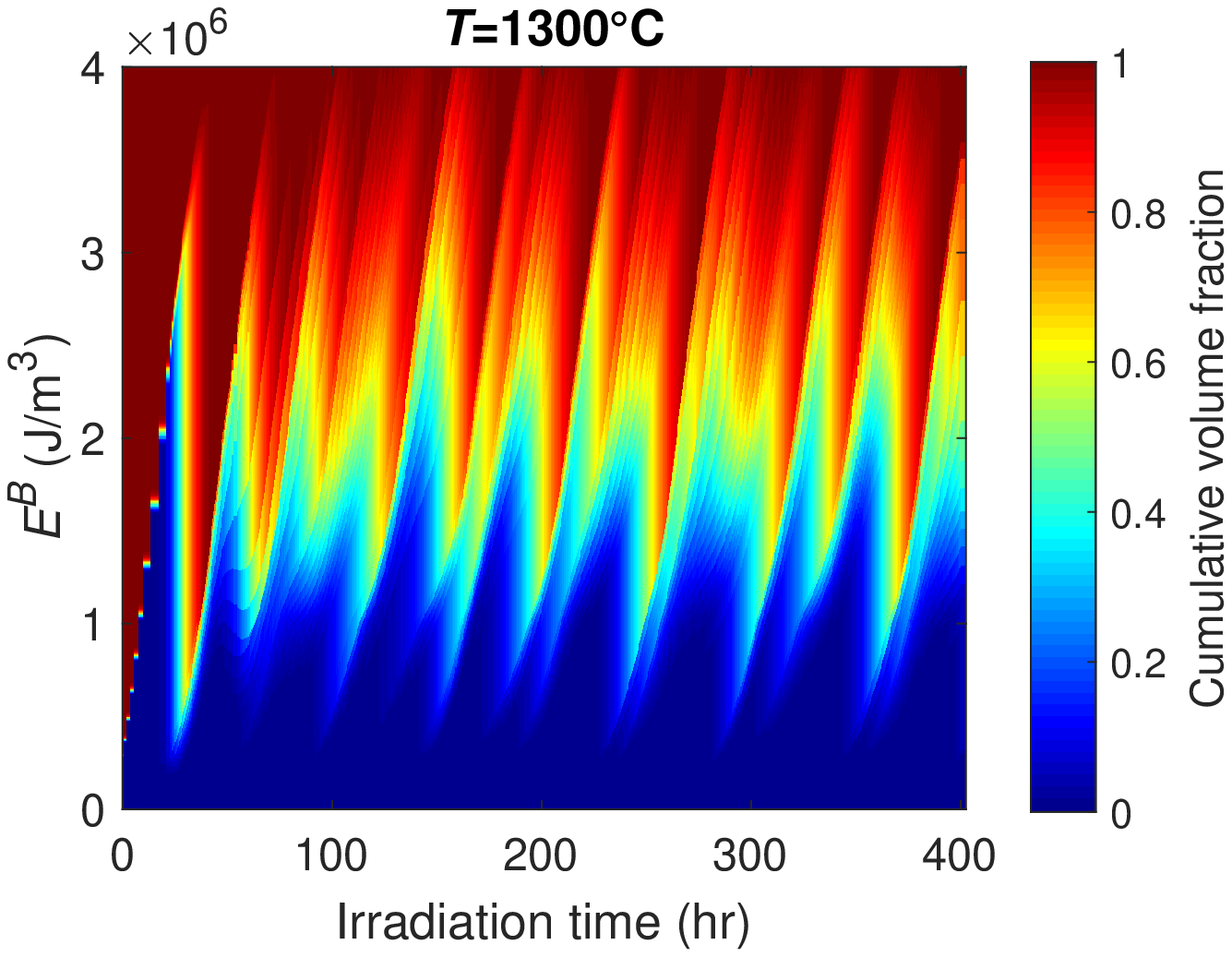}} \hspace{0.1cm}
%\subfloat[]{\includegraphics[height=0.22\textheight]{ImprUnf4_Heatmap_cumsum1400.eps}} 
\caption[]{Evolution of the grain energy distributions at (a) 1000 \textdegree C, (b) 1100 \textdegree C and (c) 1300 \textdegree C, in terms of cumulative volume fractions; Final concentration distributions for (c) $N_{max}$=100 and (d) $N_{max}$=1000.}
\label{fig:edist}
\end{figure}

%\newpage
%\begin{thebibliography}{99}
%\printbibliography
%\end{thebibliography}

\end{document}